\documentclass[superscriptaddress,showpacs,amsmath,amssymb,floatfix]{revtex4}
\usepackage{graphicx}% Include figure files
\usepackage{bm}% bold math
\begin{document}

\title{
Persistence in nonequilibrium surface growth}

\author{ M. Constantin}
\affiliation{
Condensed Matter Theory Center, 
Department of Physics, University of Maryland, College Park, Maryland 
20742-4111, USA}
\affiliation{Materials Research Science and Engineering Center, 
Department of Physics, University of Maryland, College Park, Maryland 20742-4111, USA}
\author{ C. Dasgupta}
\affiliation{
Condensed Matter Theory Center, 
Department of Physics, University of Maryland, College Park, Maryland 
20742-4111, USA}
\affiliation{Department of Physics, Indian Institute
of Science, Bangalore 560012, India}
\author{ P. Punyindu Chatraphorn}
\affiliation{
Condensed Matter Theory Center, 
Department of Physics, University of Maryland, College Park, Maryland 
20742-4111, USA}
\affiliation{ Department of Physics, Faculty of Science, Chulalongkorn 
University, Bangkok, Thailand}
\author{Satya N. Majumdar}
\affiliation{Laboratoire de Physique Quantique, UMR C5626, Universite Paul
Sabatier, 31062 Toulouse Cedex, France}
\affiliation{Laboratoire de Physique Theorique et Modeles Statistiques,
Universite Paris-Sud,
Bat. 100, 91405 ORSAY Cedex, France}
\author{ S. \surname{Das Sarma}}
\affiliation{
Condensed Matter Theory Center, 
Department of Physics, University of Maryland, College Park, Maryland 
20742-4111, USA}

\begin{abstract}
Persistence probabilities of the interface height in (1+1)-- 
and (2+1)--dimensional atomistic, solid--on--solid, stochastic 
models of surface growth are studied using kinetic Monte 
Carlo simulations, with emphasis on models that belong to 
the molecular beam epitaxy (MBE) universality class. Both 
the initial transient and the long--time steady--state regimes 
are investigated. We show that for growth models in the
MBE universality class, the nonlinearity of the underlying 
dynamical equation is clearly reflected in the difference 
between the measured values of the positive and negative 
persistence exponents in both transient and steady--state 
regimes. For the MBE universality class, the positive and 
negative persistence exponents in the steady--state are 
found to be $\theta^S_{+} = 0.66 \pm 0.02$ and 
$\theta^S_{-} = 0.78 \pm 0.02$, respectively, in (1+1) 
dimensions, and $\theta^S_{+} = 0.76 \pm 0.02$ and 
$\theta^S_{-} =0.85 \pm 0.02$, respectively, in (2+1) 
dimensions. The noise reduction technique is applied on 
some of the (1+1)--dimensional models in order to obtain 
accurate values of the persistence exponents. We show 
analytically that a relation between the steady--state 
persistence exponent and the dynamic growth exponent, 
found earlier to be valid for linear models, should be 
satisfied by the smaller of the two steady-state persistence 
exponents in the nonlinear models. Our numerical results for 
the persistence exponents are consistent with this prediction. 
We also find that the steady--state persistence exponents can be 
obtained from simulations over times that are much shorter than 
that required for the interface to reach the steady state. 
The dependence of the persistence probability on the system size 
and the sampling time is shown to be described by a simple scaling form.
\end{abstract}

\maketitle

%===================================================================
\section{
Introduction
}
\label{intro}
Nonequilibrium surface growth and interface dynamics represent an 
area of research that has received much attention in the last two 
decades~\cite{books}. A large number of discrete atomistic growth 
models~\cite{F,DT,WV,KD,DGK,KK_1,KPK} and stochastic growth 
equations~\cite{MH,EW,KPZ,LD,vill,chandan} have been 
found~\cite{krugrev} to exhibit {\it generic scale invariance} 
characterized by power law behavior of several quantities of 
interest, such as the interface width as a function of time 
(measured in units of deposited layers) and space- and 
time--dependent correlation functions of the interface height.
Much effort has been devoted to the classification of 
growth models and equations into different universality 
classes characterized by the values of the exponents that 
describe the dynamic scaling behavior implied by these 
power laws. A variety of experimental studies~\cite{krugrev,expt}
have confirmed the occurrence of dynamic scaling in nonequilibrium
epitaxial growth. Among the various experimental methods of 
surface growth, molecular beam epitaxy (MBE) is especially 
important because it plays a crucial role in the fabrication of
smooth semiconductor films required in technological applications. 
Under usual MBE growth conditions, desorption from the film 
surface is negligible and the formation of bulk vacancies and 
overhangs is strongly suppressed. It is generally believed that 
nonequilibrium surface growth under these conditions is 
well--described by a {\it conserved} nonlinear Langevin--type 
equation~\cite{LD,vill,chandan} and related atomistic 
models~\cite{DT,KD,KPK,chandan} that form the so--called 
``MBE universality class''.

Surface growth is an example of a general class of problems involving
the dynamics of non--Markovian, spatially extended, stochastic systems.
In recent years, the concept of {\it persistence}~\cite{pers_review} 
has proven to be very useful in analyzing the dynamical behavior 
of such systems \cite{diffusion,IP1,IP2,I3,Cueille1,Krug1,Krug2}.
Loosely speaking, a stochastic variable is {\it persistent} 
if it has a tendency to maintain its initial characteristics 
over a long period of time. The {\it persistence probability} 
$P(t)$ is typically defined as the probability that a 
characteristic feature (e.g. the sign) of a stochastic variable 
does not change at all over a certain period of time $t$. Although the 
mathematical concept of persistence was introduced a long time 
ago in the context of the ``zero--crossing problem'' in Gaussian 
stationary processes \cite{slepian}, it is only very recently 
that this concept has received  attention in describing the 
statistics of first passage events in a variety of spatially 
extended nonequilibrium systems. Examples of such applications 
of the concept of persistence range from the fundamental 
classical diffusion equation \cite{diffusion} to the 
zero temperature Glauber dynamics of the ferromagnetic Ising and 
$q$--state Potts models \cite{IP1,IP2,I3,Dornic1} and  phase ordering 
kinetics \cite{Cueille1}. Recently, a generalization of the 
persistence concept (probability of persistent large deviations) 
has been introduced \cite{Dornic1}. A closely related idea, 
that of sign--time distribution, was developed in Ref.~\cite{zoltan}.
An increasing number of experimental results are also available 
for persistence in systems such as coalescence of droplets 
\cite{droplet}, coarsening of two--dimensional soap froth 
\cite{soap}, twisted nematic liquid crystal \cite{liqcryst}, and
nuclear spin distribution in laser polarized $Xe^{129}$ gas 
\cite{Xe}.

Recent work of Krug and collaborators \cite{Krug1,Krug2} has 
extended the persistence concept to the first--passage statistics of 
fluctuating interfaces. Persistence in the dynamics of 
fluctuating interfaces is of crucial importance in ultra--small 
scale solid--state devices. As the technology advances into the 
nanometric regime, questions such as how long a particular 
perturbation that appears in an evolving interface persists 
in time and what is the average time required for a structure to 
first fluctuate into an unstable configuration become important. The
persistence probability can provide quantitative predictions on such
questions. Recent experiments~\cite{Dan,alex,family} have demonstrated
the usefulness of the concept of persistence in the characterization 
of the equilibrium fluctuations of steps on a vicinal surface. 
Analysis of experimental data on step fluctuations on Al/Si(111) 
\cite{Dan,family} and Ag(111) \cite{alex,family} surfaces has shown
that the long--time behavior of the persistence probability and 
the probability of persistent large deviations in these systems 
agrees quantitatively with the corresponding theoretical 
predictions. These results show that the persistence probability 
and related quantities are particularly relevant for describing and 
understanding the long--time dynamics of interface fluctuations.

In the context of surface growth and fluctuations, the persistence 
probability $P(t_0,t_0+t)$ may be defined as the probability that 
starting from an initial time $t_{0}$, the interfacial height 
$h({\bf r},t^\prime)$ at spatial position $\bf r$ does not 
return to its original value at any point in the time interval 
between $t_0$ and $t_0+t$. This probability is clearly the 
sum of the probabilities of the height $h({\bf r},t^\prime)$ 
always remaining above (the {\it positive} persistence 
probability $P_{+}$) and always remaining below (the 
{\it negative} persistence probability $P_{-}$) its 
specific initial value $h({\bf r},t_0)$ for all 
$t_0 < t^\prime \le t_0+t$. This concept quantifies the 
tendency of a stochastic field (in our case the interface height) 
to persistently conserve a specific feature (the sign of the 
interfacial height fluctuations). The persistence probability
$P(t_0,t_0+t)$ would, in general, depend on both $t_0$ and $t$. 
In the early stage of the growth process starting from a flat 
interface (transient regime), the interface gradually develops 
dynamical roughness \cite{books} due to the effect of fluctuations in 
the beam intensity. In this regime, the choice of the initial 
time $t_0$ is clearly important: it determines the degree of
roughness of the configuration from which the interface evolves.
At long times, the growing interface enters into a new evolution 
stage, called the steady--state regime, characterized by fully 
developed roughness that does not increase further in time. In 
this regime, the choice of $t_0$ is expected to be unimportant. 

The work of Krug {\it et al.}~\cite{Krug1} shows that for a class 
of {\it linear} Langevin--type equations for surface growth and 
atomistic models belonging in the same dynamical universality 
class as these equations, the persistence probability decays as 
a power law in time for long times in both transient and steady--state 
regimes. These power laws define the positive and negative 
persistence exponents, $\theta_{\pm}^T$ and $\theta_{\pm}^S$, 
for positive and negative persistence in the transient and 
steady--state regimes, respectively. The $h \to -h$ symmetry of the
linear growth equations implies that $\theta_{+}^T = \theta_{-}^T$ and
$\theta_{+}^S = \theta_{-}^S$ in these systems. In Ref.~\onlinecite{Krug1}, 
it was pointed out that the persistence exponent in the steady 
state of these linear models is related to the dynamic scaling 
exponent $\beta$, that describes the growth of the interface width 
$W$ as a function of time $t$ in the transient regime ($W \propto
t^\beta$), through the relation $\theta_{+}^S = \theta_{-}^S
= 1-\beta$. The validity of this relation was confirmed by numerical
simulations. Since the exponent $\beta$ is the same for all models in the
same dynamical universality class, this result implies that the persistence
exponent in the steady--state regime of these {\em linear} models is also
universal. Numerical results for the persistence exponent in the
transient regime, for which no analytic predictions are available, 
also indicate a similar universality. Kallabis and Krug \cite{Krug2} 
carried out a similar calculation for (1+1)--dimensional
Kardar--Parisi--Zhang (KPZ)~\cite{KPZ} interfaces. They found that the 
nonlinearity in the KPZ equation that breaks the $h \to -h$ 
symmetry is reflected in {\it different} values of the positive 
and negative persistence exponents, $\theta_{+}^T$ and $\theta_{-}^T$, 
in the transient regime. The values of the steady--state persistence
exponents $\theta_{+}^S$ and $\theta_{-}^S$ were found to be equal to each
other, and equal to $1-\beta$ within the accuracy of the numerical results.    
This is expected because the $h \to -h$ symmetry is dynamically 
restored in the steady state of the (1+1)--dimensional KPZ
equation. This is, however, a specific feature of the
(1+1)--dimensional KPZ model, which for nongeneric reasons, turns out
to be up--down symmetric in the steady state. Nonlinear surface growth
models (e.g. the higher dimensional KPZ model, the nonlinear MBE
growth model) are generically expected to have different values of
$\theta_{\pm}$ in both transient and steady--state regimes.

In this paper, we present the results of a detailed numerical 
study of the persistence behavior of several atomistic, 
solid--on--solid (SOS) models of surface growth in (1+1) and 
(2+1) dimensions. While we concentrate on models in the MBE 
universality class, results for a few other models, some of 
which have been studied in Refs.~\onlinecite{Krug1} and 
\onlinecite{Krug2} are also presented for completeness. 
The highly non--trivial nature of the persistence probability, 
in spite of a deceptive simplicity of the defining concept, 
arises from the complex temporal non--locality (``memory'') 
inherent in its definition. In fact, there are very few 
stochastic problems where an analytical solution for the 
persistence probability has been achieved. These include 
the classical Brownian motion \cite{Feller}, the random acceleration 
problem \cite{Sinai} and the one dimensional Ising and $q$-state Potts
models \cite{IP2}. In general, the highly nonlocal
nature of the temporal correlations in a non--Markovian stochastic
process makes it extremely difficult to obtain exact results for the
persistence probability even for seemingly simple stochastic
processes. Even for the simple diffusion equation, the 
persistence exponent is known only numerically, or within an 
independent interval approximation \cite{diffusion} or series 
expansion approach \cite{SNM}. 
However, it is fairly straightforward in most cases to 
directly simulate the persistence probability to obtain its 
stationary power law behavior at large times, and thus to 
numerically obtain the approximate value of the persistence 
exponent. For this reason, we use stochastic (Monte Carlo) 
simulations of the atomistic growth models to study their temporal 
persistence behavior in the transient and steady--state regimes. 
These models are defined in terms of random deposition and specific 
cellular--automaton--type local diffusion or relaxation rules. 
Some of these models 
are of the ``limited--mobility'' type in the sense that the surface 
diffusion rules or local restrictions limit the characteristic 
length over which a deposited particle can diffuse to just one 
or a few lattice spacings. The models in the MBE universality 
class considered in our study are: the Das Sarma--Tamborenea 
model~\cite{DT}, the Wolf--Villain model~\cite{WV,note}, the 
Kim--Das Sarma model~\cite{KD} and its ``controlled'' version~
\cite{chandan}, and the restricted solid--on--solid (RSOS) model 
of Kim {\it at al.}~\cite{KPK}. We also present results for the 
Family model~\cite{F} that is known to belong to the 
Edwards--Wilkinson~\cite{EW} universality class and the restricted
solid--on--solid (RSOS) model of Kim and Kosterlitz~\cite{KK_1} 
that is in the KPZ universality class.

The main objective of our study is to examine the effects of 
the nonlinearity in the MBE growth equation~\cite{LD,vill} on 
the persistence behavior. Unlike the (1+1)--dimensional KPZ 
equation, the nonlinearity in the MBE growth equation persists 
in the steady state in the sense that the height profile exhibits 
a clear asymmetry between the positive and negative directions 
(above and below the average height). Therefore, the positive
and negative persistence exponents $\theta_{+}^S$ and 
$\theta_{-}^S$ are expected to have different values in these 
models. If this is the case, then the relation between the 
steady--state persistence exponent and the dynamic scaling 
exponent $\beta$ found in linear models can not be valid
for both $\theta_{+}^S$ and $\theta_{-}^S$, indicating that 
at least one of these exponents is a new, nontrivial one not 
related to the usual dynamic scaling exponents. The values of 
$\theta_{+}^S$ and $\theta_{-}^S$ and their relation to $\beta$, 
as well as the values of the transient persistence exponents 
$\theta_{+}^T$ and $\theta_{-}^T$ are the primary questions 
addressed in our study. We also investigate the universality of 
these exponents by measuring them for several models that
are known to belong in the same universality class as far as their
dynamic scaling behavior is concerned. To obtain accurate values of the 
exponents, the ``noise reduction'' technique~\cite{nrt3} 
is employed in some of the simulations of (1+1)--dimensional 
models. We also address some questions related to the methodology 
of calculating persistence exponents from simulations. Since the 
value of the dynamical exponent $z$ is relatively large for models 
in the MBE universality class, the time required for reaching 
the steady state grows quickly as the sample size $L$ is increased
($t_{sat} \propto L^z$). As a result, it is difficult to reach 
the steady state in simulations for large $L$. It is, therefore, 
useful to find out whether the value of the steady--state persistence 
exponents can be extracted from calculations of $P(t_0,t_0+t)$ 
with $t_0 \ll t_{sat} \propto L^z$. Another issue in this context 
involves the effects of the finiteness of the sample size $L$ 
and the sampling time $\delta t$ (the time interval between two
successive measurements of the height profile) on the calculated
persistence probability. An understanding of these effects is needed for
extracting reliable values of the persistence exponents from simulations
that {\it always} involve finite values of $L$ and $\delta
t$. Understanding the effects of $L$ and $\delta t$ on the persistence
analysis is not only important for our simulations, but is also
important in the experimental measurements of persistence which
invariably involve finite system size and sampling time.
 
The main results of our study are as follows. We find that the 
positive and negative steady--state persistence exponents for 
growth models in the MBE universality class are indeed different 
from each other, reflecting the asymmetry of the interface
arising from the presence of nonlinearities in the underlying growth 
equation. Our results for these exponent values are:
$\theta^S_{+} = 0.66 \pm 0.02$ and $\theta^S_{-} = 0.78 \pm 0.02$, 
respectively, in (1+1) dimensions; $\theta^S_{+} = 0.76 \pm 0.02$ and 
$\theta^S_{-} =0.85 \pm 0.02$ in (2+1) dimensions. The values of the
positive and negative persistence exponents for different models are 
clearly correlated with the asymmetry of the ``above'' and ''below''
(defined relative to the mean interface height) portions of the interface.
We show analytically that the {\it smaller} one of the two steady-state
persistence exponents should be equal to $(1-\beta)$. Thus, 
the relation $\theta=1-\beta$ derived in Ref.~\cite{Krug1} 
for linear surface growth models is expected to be
satisfied by $\theta_{+}^{S}$ for the nonlinear models considered here.
Our numerical results are consistent with this expectation: we find that
the positive persistence exponent is indeed 
close to $(1-\beta)$, while the negative
one is significantly higher. 
%\approx 1-\beta$ for
%the positive steady--state persistence exponent in the nonlinear MBE
%growth model, but $not$ for $\theta_{-}^S$. 
Similar asymmetry is 
found for the persistence exponents in the transient regime with
$\theta_{+}^{T}< \theta_{-}^{T}$ in MBE growth. Within the
uncertainties in the numerically determined values of the exponents, 
they are universal in the sense that different models in the same 
dynamic universality class yield very similar values for these 
exponents. For the models in the Edwards--Wilkinson and KPZ 
universality classes, we find results in agreement with those of
earlier studies~\cite{Krug1,Krug2}.

Our simulations also reveal that a measurement of the steady--state 
persistence exponents is possible from simulations in which the 
initial time $t_0$ is much smaller than the time ($\sim L^z$) 
required for the interfacial roughness to saturate. A similar 
result was reported in Ref.~\onlinecite{Krug1} where it was 
found that the steady--state persistence exponent may be obtained from a
calculation of $P(t_0,t_0+t)$ with $t \ll t_0 \ll L^z$. We find that the
restriction $t \ll t_0$ is not necessary for seeing a power law behavior of 
$P_{\pm}(t_0,t_0+t)$ -- a power law with the steady--state exponents is found
even if $t$ is close to or somewhat larger than $t_0$. We exploit this 
finding in some of our persistence simulations for (2+1)--dimensional 
growth models which are more relevant to experiments. These results, 
however, also imply that it would be extremely difficult to 
measure the transient persistence exponents from real surface growth
experiments. Finally, we show that the dependence of the steady--state
persistence probability on the sample size $L$ and the sampling 
time $\delta t$ is described by a simple scaling function of the 
variables $t/L^z$ and $\delta t/L^z$. This scaling description 
is similar to that found recently~\cite{p0} for a different 
``persistence probability'', the survival probability, that 
measures the probability of the height not returning to its 
{\it average value} (rather than the initial value) over a certain 
period of time. Although the ``persistence'' and the ``survival''
\cite{p0} probability seem to be qualitatively similar in their
definitions, the two are mathematically quite unrelated, and in fact,
no exponent can be defined for the survival probability. In this paper
we only discuss the persistence probability and the persistence
exponent for surface growth processes.

The rest of the paper is organized as follows: in Sec. 
\ref{A}, we briefly discuss the main universality classes and their 
corresponding dynamic equations and scaling exponents relevant for 
surface growth phenomena. Section \ref{B} contains a short 
overview of the persistence probability concept from the interface 
fluctuations perspective. The discrete stochastic SOS growth 
models considered in our study are described in Sec. 
\ref{models}. In addition, we briefly describe in this 
section the noise reduction technique which is employed in 
some of our simulations. Section \ref{sim} contains a detailed 
description of our main results: in Sec. \ref{sim1}, 
our (1+1)--dimensional simulation results for the transient 
and steady--state persistence exponents are presented, focusing 
mostly on models described by the nonlinear MBE dynamical 
equation. Section \ref{analytic} contains the analytic derivation of a
relation between the smaller steady-state persistence exponent and
the dynamic growth exponent. 
In Sec. \ref{sim-steady}, we introduce an alternative 
approach for measuring the steady--state persistence exponents, 
using a relatively short ``equilibration time'' that is much 
shorter than the time required for reaching the true steady state.
Section \ref{sim2} contains the results of our (2+1)--dimensional 
persistence calculation for a selection of linear and nonlinear 
models. Simulation results that establish a scaling form of the 
dependence of the persistence probability on the sample size 
and the sampling time are presented in Sec. \ref{fss}. The 
final Sec. \ref{conclude} contains a summary of our main 
results and a few concluding remarks.

%================================================================
\section{
Stochastic growth 
equations and persistence probabilities
}
\label{theo}
\subsection{Growth equations and dynamic scaling}
\label{A}
The dynamic scaling behavior of stochastic growth equations 
may be classified into several universality classes.
Each universality class is characterized by a set of 
scaling exponents \cite{books} which depend on the 
dimensionality of the problem. These exponents are 
$(\alpha, \beta, z)$, where $\alpha$ is the roughness 
exponent describing the dependence of the amplitude of 
height fluctuations in the steady--state regime 
($t \gg L^{z}$) on the sample size $L$, $\beta$ is 
the growth exponent that describes the initial power law 
growth of the interface width in the transient regime 
($1 \ll t \ll L^{z}$), and $z$ is the dynamical 
exponent related to the system size dependence of the 
time at which the interface width reaches saturation. Note 
that $z=\frac{\alpha}{\beta}$ for all the models considered 
in this paper. To describe the interface evolution we use 
the single--valued function, $h({\bf r},t)$, which represents 
the height of the growing sample at position ${\bf r}$ and 
deposition time $t$. The interfacial height fluctuations 
are described by the root--mean--squared height deviation 
(or interface width) which is a function of the substrate 
size $L$ and deposition time $t$:
\begin{equation}
\label{w}
W(L,t) = \langle ( h({\bf r},t) - \bar h(t) )^{2} \rangle^{1/2} ,
\end{equation}
\noindent where $\bar h(t)$ is the average sample thickness. The width
$W(L,t)$ scales as $W(L,t) \propto t^{\beta}$ for $t \ll L^{z}$ 
and $W(L,t) \propto L^{\alpha}$ for $t \gg L^{z}$ 
\cite{Fam_scal}, $L^{z}$ being the equilibration time of the 
interface, when its stationary roughness is fully developed.

Since it is convenient to write the evolution equations in terms 
of the deviation of the height from its spatial average 
value, $h({\bf r},t)-\bar h(t)$, from now on we will denote 
by $h({\bf r},t)$ the interface height fluctuation measured 
from the average height. Extensive studies of dynamic scaling in 
kinetic surface roughening (for an extended review see 
Ref.~\cite{krugrev}) have revealed the existence of (at least) 
four universality classes that are described, in the long 
wavelength limit, by the following continuum equations and sets 
of scaling exponents ($\alpha,\beta,z$), shown for the 
1+1(2+1)--dimensional cases, respectively:

\noindent(1) The Edwards--Wilkinson (EW) second order linear equation: 
$\frac{1}{2},~\frac{1}{4},~2$ $(0~(\log), 0~(\log), 2)$
\begin{equation}
\label{EW_eq}
\frac {\partial h({\bf r}, t)} {\partial t} = \nu_{2} 
\nabla^{2} h({\bf r},t) + \eta({\bf r}, t),
\end{equation}
(2) The KPZ second order nonlinear equation: 
$\frac{1}{2},~\frac{1}{3},~\frac{3}{2}$ $(\simeq 0.4,~\simeq 0.24,~\simeq 1.67)$
\begin{equation}
\label{KPZ_eq}
\frac {\partial h({\bf r}, t)} {\partial t} = 
\nu_{2} \nabla^{2} h({\bf r},t) +
 \lambda_{2} |{\bf \nabla} h({\bf r},t) 
|^{2} + \eta({\bf r}, t),
\end{equation}
(3) The Mullins--Herring (MH) fourth order linear equation: 
$\frac{3}{2},~\frac{3}{8},~4$ $(1,~\frac{1}{4},~4)$
\begin{equation}
\label{MH_eq}
\frac {\partial h({\bf r}, t)} {\partial t} = -\nu_{4} 
\nabla^{4} h({\bf r},t) + \eta({\bf r}, t),
\end{equation}

\noindent and

\noindent(4) The MBE fourth order nonlinear equation: 
$\simeq 1,~\simeq \frac{1}{3},~\simeq 3$ $(\simeq \frac{2}{3},~\simeq 
\frac{1}{5},~\simeq \frac{10}{3})$
\begin{equation}
\label{MBE_eq}
\frac {\partial h({\bf r}, t)} {\partial t} = -\nu_{4} \nabla^{4} h({\bf r},t) +
\lambda_{22} \nabla^{2} |( {\bf \nabla} h({\bf r},t)|^{2} + 
\eta({\bf r}, t),
\end{equation}
\noindent where $\nu_{i}$ (i=2, 4) and $\lambda_{j}$ (j=2, 22) 
are constant. The quantity $\eta({\bf r}, t)$ represents the noise 
term which accounts for the random fluctuations in the deposition 
rate. We assume that the noise has Gaussian distribution with 
zero mean and correlator

\begin{equation}
\label{noise}
\langle \eta ({\bf r}_{1}, t_{1}) \eta( {\bf r}_{2}, t_{2}) 
\rangle = D\delta({\bf r}_{1}- 
{\bf r}_{2})\delta(t_{1}-t_{2}),
\end{equation}
\noindent D being a constant related to the strength of the bare
noise. Note that we do not include the (trivial) constant external
deposition flux term in the continuum growth equations since that is
easily eliminated by assuming that the height fluctuation $h$ is
always measured with respect to the average interface which is growing
at a constant rate. 

The concepts of universality classes and scaling exponents have 
been widely used in the literature to analyze the kinetics of 
surface growth and fluctuations. Our study based on persistence 
probabilities is motivated by the possibility that the concept of 
persistence may provide an additional (and complementary) tool 
to analyze the surface growth kinetics. It addresses fundamental 
questions such as: is persistence an independent 
(and new) conceptual tool for studying surface fluctuations 
or essentially equivalent (or perhaps complementary) to 
dynamic scaling? and does persistence lead to the definition 
of new universality classes on the basis of the values of the
persistence exponent? To answer these questions, we consider,
for each of the four universality classes mentioned above 
(i.e. Eqs.~(\ref{EW_eq})--(\ref{MBE_eq})), at least one growth 
model and investigate how the associated persistence exponents 
are related to the dynamic scaling exponents mentioned above.

%===============================================================
\subsection{Transient and steady--state persistence probabilities}
\label{B}

Our goal is to calculate the positive and negative persistence 
probabilities ($P_{\pm}(t_{0},t_{0}+t)$) for a growing 
(fluctuating) interface in the transient and steady--state 
regimes. Here $t_{0}$ is the initial time, and we are 
interested in evaluating the probability of the height 
at a fixed position remaining persistently above ($P_{+}$) 
or below ($P_{-}$) its initial value (i.e. its value at $t_0$ by
definition) during the time 
period between $t_0$ and $t_0+t$. If one considers the 
special case $t_{0}=0$, when the interface is completely 
flat, then the quantity of interest is the probability that 
the interfacial height (measured from its spatial average) 
does not return to its initial zero value up to time $t$. 
This case is known as the {\it transient} (T) regime. For 
values of $t$ that are small compared to the time scale for 
saturation of the interface width ($t_{sat}(L) \propto L^z$), 
the persistence probabilities in this regime are expected 
to exhibit a power law decay in time:

\begin{equation}
\label{P+-} P_{\pm}^{T}(0,t) \propto \left( 
\frac{1}{t}\right)^{\theta_{\pm}^{T}},
\end{equation}
where $\theta_{\pm}^{T}$ are called the transient 
positive and negative persistence exponents. In the particular 
case of {\it linear} continuum growth equations, these exponents 
are equal because the symmetry under a change of sign 
of $h({\bf r}, t)$ remains valid at all stages of the growth 
process. However, in the case of dynamics governed by nonlinear 
continuum equations, the lack of this ``up--down'' interfacial 
symmetry implies that $P_{+}$ and $P_{-}$ (and therefore, 
the exponents $\theta_{+}^T$ and $\theta_{-}^T$) would, 
in general, be different from each other. No universal 
relationship between the transient positive and negative persistence 
exponents and the dynamic scaling exponents is known to 
exist for any one of the four universality classes mentioned 
above.

On the other hand, if one considers $t_{0}$ larger than $t_{sat}(L)$, 
then the quantity of interest is the probability that the 
interfacial height at a fixed position does not return to 
its specific value at initial time $t_0$ during the subsequent 
time interval between $t_0$ and $t_0+t$. Instead of being flat, 
the interface morphology at time $t_{0}$ has completely 
developed roughness, which produces persistence exponents 
that are different from the transient exponents defined
earlier. This case is known as the {\it steady--state} (S) 
regime. If $t \ll L^z$, one expects to obtain in this regime 
the steady--state persistence probability with a power law decay 
in time \cite{Krug1}

\begin{equation}
\label{P_steady} P_{\pm}^{S}(t_{0},t_{0} + t) \propto \left( 
\frac{1}{t}\right)^{\theta_{\pm}^{S}}.
\end{equation}
where $\theta_{\pm}^{S}$ are the steady--state positive and 
negative persistence exponents. It has been pointed out by 
Krug {\it et al.} \cite{Krug1} that for systems described 
by {\it linear} Langevin equation, the steady--state
persistence exponents are related to the dynamic scaling 
exponent $\beta$ in the following way:
\begin{equation}
\label{theta_S}
\theta_{+}^{S}\equiv \theta_{-}^{S}=1-\beta.
\end{equation}
The exponent $\beta$ is well known for linear Langevin 
equations for surface growth dynamics, and is given 
in $d$--dimensions by $\beta=(1-d/z)/2$  for nonconserved 
white noise (Eq.~(\ref{noise})), where $z$, the dynamical exponent, is
here precisely equal to the power of the gradient operator entering
the linear continuum dynamical growth equation [i.e. $z=2$ in
Eq.~(\ref{EW_eq}); $z=4$ in Eq.~(\ref{MH_eq})]. The relation 
defined by Eq.~(\ref{theta_S}) holds true for the Langevin equations of
Eqs.~(\ref{EW_eq}) and (\ref{MH_eq}), which are obviously linear, 
as well as for the special case of the (1+1)--dimensional KPZ 
equation of Eq.~(\ref{KPZ_eq})\cite{Krug2}, which, despite its 
nonlinearity, behaves as the linear EW equation in the 
steady state. Since the positive and negative exponents are 
expected to be different for general nonlinear Langevin equations, 
the relation of Eq.~(\ref{theta_S}) can not be valid for both 
$\theta_{+}^S$ and $\theta_{-}^S$ in systems described by such 
nonlinear equations. Therefore, at least one (or perhaps both) 
of these two persistence exponents must be non--trivial in the 
sense that it is not related to the usual dynamic scaling 
exponents. For this reason we pay particular attention to 
the MBE nonlinear equation and investigate whether its 
persistence exponents can be related to the dynamic scaling exponents.
 
%===================================================================
\section{Atomistic growth models}
\label{models} In this paper, we use different atomistic limited--mobility 
growth models for simulating surface growth processes. In these 
models, the substrate consists of a collection of lattice sites 
labeled by the index $j$ ($j=1,2,\ldots,L^d$) and the height 
variables {$h(x_{j})$} take integral values. The term
``limited--mobility'' is meant to imply that in these models,
each adatom is characterized by a finite diffusion length 
which is taken to be one lattice spacing in most of the 
models we consider here. Thus, a deposited atom can explore 
only a few neighboring lattice sites according to a set of 
specific mobility rules before being incorporated into the 
growing film. The solid--on--solid constraint is imposed 
in all these models, so that defects such as overhangs 
and bulk vacancies are not allowed. In most of the models 
considered in this work, the possibility of desorption is 
neglected, thereby making the models ``conserved'' in the 
sense that all deposited atoms are incorporated in the film; the noise
(given by Eq.~(\ref{noise})) is of course nonconserved since the
system is open to the deposition flux.

The deposition process is described by a few simple rules in 
these models. An atomic beam drops atoms on the substrate in 
a random manner. Once a lattice site on the substrate is randomly 
chosen, the diffusion rules of the model are applied to the atom 
dropped at the chosen site to determine where it should be 
incorporated. The allocated site is then instantaneously 
filled by the adatom. We consider both (1+1)-- and (2+1)--
dimensional models (one or two spatial dimensions and one 
temporal dimension) defined on substrates of length $L$ in 
units of the lattice spacing. The deposition rate is taken to be 
constant and equal to $L^d$ particles per unit time in our 
simulations of the Family (F), larger curvature (LC), 
Das Sarma--Tamborenea (DT), Wolf--Villain (WV) and controlled 
Kim--Das Sarma (CKD) models (see below). In these simulations, 
one complete layer is grown in each unit of time. In the 
RSOS Kim--Kosterlitz (KK) and Kim--Park--Kim (KPK) models 
described below, the diffusion rules are replaced  by a set of
local restrictions on nearest neighbors height differences, which 
have to be satisfied after the deposition. The randomly 
chosen deposition site is rejected (the atom is not deposited)
if these restrictions are not satisfied. As a consequence, 
the number of deposition attempts does not coincide with the 
number of successful depositions in the KK model, although they 
are linearly related. 

All conserved growth models satisfy the conservation law 
\begin{equation}
\label{current} \frac {\partial h({\bf r},t)}{\partial t}= - {\bf
\nabla} \cdot {\bf j}({\bf r},t) + \eta({\bf r},t),
\end{equation}
where $\bf j$ is the surface current and 
$\eta$ is the noise term. Using different expressions, dictated 
primarily by symmetry considerations, for the current $\bf j$,
one can obtain all the conserved Langevin equations discussed in
Sec. \ref{A}. The atomistic growth models considered in our work
provide discrete realizations of these continuum growth equations.

It is known that some of the discrete growth models we study here
have complicated transient behavior \cite{WVuniv,DT2d}. For 
this reason, obtaining the dynamic scaling exponents that 
show the true universality classes of these models is often 
quite difficult. To make this task easier, the noise reduction 
technique \cite{nrt1,nrt2} was introduced in simulations of 
such models. It has been shown \cite{nrt3} that this technique 
helps in suppressing high steps in the models and reduces the 
corrections in the scaling behavior, so that the true asymptotic 
universality classes of the growth models can be seen in 
simulations that cover a relatively short time. This makes it 
interesting to examine whether the persistence probabilities in these 
discrete models also exhibit similar transient behavior, and whether 
the noise reduction technique can help in bringing out the true persistence 
exponents of these models. To investigate this, we have applied the noise 
reduction technique to some of the discrete models studied in 
this paper.

The noise reduction technique can be easily incorporated in the 
simulation of any discrete growth model by a small modification 
in the diffusion process \cite{nrt3}. When an atom is dropped 
randomly, the regular diffusion rules for the growth model are 
applied and the final allocated site is chosen. Instead of adding the atom at 
that final site, a counter at that site is increased by one but 
the height of that site remains unchanged. When the counter of a 
lattice site increases to the value of a pre--determined noise 
reduction factor, denoted by $m$, the height at that lattice site 
is increased by one and the counter of that site is reset back to 
zero. The value of the noise reduction factor $m$ should be chosen 
carefully. If $m$ is too small, the suppression of the noise 
effect is not enough and the true universality class is not seen. 
However, is $m$ is too large, the kinetically rough growth 
becomes layer--by--layer growth \cite{layer} and the universality 
class of the model cannot be determined.

The atomistic models considered in our work are defined below.
%=====================================================================

(i) {\bf Family model}:
%\label{F} 
The Family (F) model \cite{F} is an extensively studied SOS 
discrete stochastic model, rigorously known to belong to the same 
dynamical universality class as the EW equation. 
It allows the adatom to explore 
within a fixed diffusion length to find the lattice site 
with the smallest height where it gets incorporated. If the diffusion length 
is one lattice constant (this is the value used in our simulations), 
the application of this deposition rule 
to a randomly selected site $j$ involves finding the local minimum 
height value among the set: $~h(x_{j-1}),~h(x_{j})$ and 
$~h(x_{j+1})$ (in (1+1)--dimensions). The height of the site with the minimum
height is then increased by one.

(ii) {\bf Larger curvature model}:
%\label{LC} 
The Kim--Das Sarma model \cite{KD} is a more complex one
which allows the atomic surface current $\bf j$ to be written as 
a gradient of a scalar field $K$, ${\bf j} = -{\bf \nabla} K$, which can 
depend on $h$, 
$\nabla^{2}h$, $|\nabla h|^{2}$ and so on. In the particular case 
when $K=-\nabla^{2}h$, one obtains the so--called larger curvature 
(LC) model. As the name suggests, the diffusion rules applied 
to a randomly selected site $j$ allow the adatom to get incorporated at the
site in the neighborhood of site $j$ where the local 
curvature (given by $h(x_{j+1})+h(x_{j-1})-2h(x_{j})$ in 
(1+1)--dimensions) has the largest value. The LC model 
asymptotically rigorously belongs to MH universality class 
described by Eq.~(\ref{MH_eq}).

(iii) {\bf Wolf--Villain model}:
%\label{WV} 
The diffusion rules of the Wolf--Villain (WV) model \cite{WV} 
allow the adatom to diffuse to its neighboring sites in 
order to maximize its local coordination number which, for the 
(1+1)--dimensional case, varies between 1 and 3 when
the bond with the atom lying below the site under 
consideration is taken into account. In contrast 
to the F model, in this case the surface develops deep valleys 
with high steps almost perpendicular to the substrate.
For the range of times and sample sizes used in the present study, 
the WV model may be considered to belong to the MBE universality class 
\cite{WV,WVuniv} described by Eq.(\ref{MBE_eq}). However, recent 
studies \cite{Kotrla,nrt3} have shown that the asymptotic
universality class of this model in (1+1)--dimensions is the 
same as that of the EW equation. In contrast, in (2+1)--dimensions,
studies based on the noise reduction technique \cite{Patcha_new} 
have revealed that the WV model exhibits at very long times unstable (mounded) dynamic
universality which cannot really be described by any of the continuum
equations (Eqs.~(\ref{EW_eq})-(\ref{MH_eq})) given above. 
   
(iv) {\bf Das Sarma--Tamborenea model}:
%\label{DT} 
The Das Sarma--Tamborenea (DT) model \cite{DT} is characterized 
by diffusion rules that are slightly different from those in the 
WV model. In this case, the diffusing atom tries to {\it increase} 
its coordination number, not necessarily to {\it maximize} it. 
For example, if a randomly selected deposition site has its 
local coordination number equal to 1 (i.e. no lateral neighbor 
in (1+1)--dimensions), and the two neighbors of this site 
have coordination numbers equal to 2 and 3, the deposited 
atom does not necessarily move to the neighboring site with the
larger local coordination number: it moves to one of the two 
neighboring sites with equal probability (the atom would necessarily 
move to the site with coordination number 3 in the WV model).    
This minor change in the local diffusion rules actually changes 
the asymptotic universality class: the (1+1)--dimensional DT model 
belongs to the MBE universality class \cite{nrt3,Patcha_new} 
corresponding to the nonlinear continuum dynamical equation of 
Eq.~(\ref{MBE_eq}). However, the (2+1)--dimensional DT model 
asymptotically belongs to the EW universality \cite{Patcha_new} at
very long times. 

(v) {\bf Controlled Kim--Das Sarma model}:
%\label{CKD} 
The Kim--Das Sarma model mentioned above provides a discrete realization 
of the continuum equation of Eq.~(\ref{MBE_eq}) if the scalar 
field $K$ is chosen to be $K=-\nabla^{2}h + \lambda_{22}(\nabla 
h)^{2}$. However, the discrete treatment of the spatial gradients
produces strong instabilities in the growth process 
due to uncontrolled growth of isolated structures, such as 
pillars or grooves. These instabilities can be easily controlled by 
introducing higher order nonlinear terms \cite{chandan}. We call 
this new model the {\it controlled} Kim--Das Sarma (CKD) model. In 
this model, the scalar field $K$ is chosen to be $K=-\nabla^{2}h + \lambda_{22} 
f(|\nabla h|^{2})$, where the nonlinear function $f$ is given by 
\begin{equation}
\label{f}
f(|\nabla h|^{2})=\frac {1-e^{-c|\nabla h|^{2}}}{c},
\end{equation}
with $c>0$ being the control parameter. The CKD diffusion rules for 
a randomly chosen deposition site, $j$, imply the minimization of 
the scalar field $K$, using the standard discretization scheme 
for the lattice derivatives $\nabla^{2} h$ and $\nabla h$:
\begin{equation}
\label{ckd_1}
(\nabla^{2} h)_{|j} = h(x_{j+1})+h(x_{j-1})-2h(x_{j}),
\end{equation}
\begin{equation}
\label{ckd_2}
|\nabla h|^{2}_{|j} = \frac{1}{4}\left[h(x_{j+1})-h(x_{j-1})\right]^{2},
\end{equation}
in (1+1)--dimensions. By carefully choosing the values for 
$c$ and $\lambda_{22}$ \cite{chandan}, one can remove the nonlinear 
growth instabilities completely and ensure an overall behavior of 
the CKD model similar to that of the DT model. 

(vi) {\bf Kim--Kosterlitz and 
Kim--Park--Kim models}:
%\label{KK} 
For completeness, we also present in this paper the 
results for the RSOS Kim--Kosterlitz (KK) \cite{KK_1} and 
Kim--Park--Kim (KPK) \cite{KPK}
models which are known to belong asymptotically to the KPZ and MBE 
universality classes, respectively. The common feature of these 
two models is the replacement of the usual diffusion rules 
of the SOS models described above by local restrictive conditions
controlling nearest--neighbor height differences. 

In the KK model, deposition sites are randomly chosen, but the 
incorporation of the adatoms into the substrate is subject to 
a specific restriction: the deposition event occurs if and only 
if the absolute value of the height difference between the 
randomly selected deposition site, $j$, and {\it each} of its 
nearest--neighboring sites remains smaller than or equal to a 
positive integer $n$ after deposition (our simulations 
were done for $n=1$). If this strict constraint is not 
satisfied, the attempted deposition of an adatom is rejected, 
and the random selection of the deposition site is repeated 
until the deposition is successfully done. Since every attempt 
to deposit an adatom is not successful, the definition of ``time'' 
in this model is not quite the same as that in the other models 
where every deposition attempt leads to the incorporation of 
a new adatom in the growing film. In the KK model, the ``time'' is 
equivalent to the average height, which is not the same as the 
number of attempted depositions per site (these two quantities are
the same in the other models considered here). The KK model is known
to belong to the KPZ universality class, and in fact provides the most
numerically efficient and accurate method for calculating the KPZ
growth exponents.

Kim {\it et al.} \cite{KPK} discovered that a slight 
change in the algorithm for choosing the incorporation site 
transforms the KK model into a new one, the KPK model, that 
belongs to the MBE universality class. The change consists of 
extending the search for appropriate incorporation sites 
(i.e, sites where the constraint on the absolute values of 
the nearest--neighbor height differences would be satisfied 
after the incorporation of an adatom) to the neighbors of 
the originally selected deposition site $j$. If the original 
site does not satisfy the constraint, then the neighboring 
sites ($j\pm 1$ in (1+1) --dimensions) are checked, and 
an adatom is incorporated at one of these sites if the 
incorporation does not violate the constraint. Otherwise, 
the search is extended to the next--nearest--neighbors of $j$, 
and so on until a suitable incorporation site is found.
We mention that in our implementation of this process, if, 
for example, both the sites $j-k$ and $j+k$  are found to be
suitable for incorporation, then one of them is chosen randomly 
without any bias. Application of this algorithm in (2+1)--dimensions
involves extending the search for suitable incorporation sites to
those lying inside circles of increasing radii around the randomly 
selected deposition site $j$. The diffusion and incorporation rules of
the KPK model \cite{KPK} lead essentially to a conserved version of
the Kim--Kosterlitz RSOS model \cite{KK_1}, and as such the continuum
growth equation corresponding to the KPK model is the conserved KPZ
equation (with nonconserved noise), which is precisely the MBE
equation; Eq.~(\ref{MBE_eq}) is the conserved version of
Eq.~(\ref{MH_eq}) with nonconserved noise in both.

%===============================================================
\section{Simulation results and discussion}
\label{sim}
\subsection{Persistence exponents in (1+1) dimensions}
\label{sim1}

Simulations for (1+1)--dimensional discrete growth models were 
carried out for $\beta=1/4$, $3/8$ and $1/3$. The value 
$\beta=1/4$ corresponds to the F model that has a relatively 
small equilibration time (of the order of $L^{2}$). The 
remaining conservative models, characterized by $\beta=3/8$ 
(LC) and $\simeq 1/3$ (WV, DT, CKD and KPK), have a much 
slower dynamics (with $z$ values 4 or 3). So their corresponding 
equilibration time intervals, required for the interface 
roughness to reach saturation, are of the order of $L^{4}$ 
and $L^{3}$, respectively. For this reason, the largest values of $L$ 
for which the steady state could be reached in reasonable 
simulation time are considerably shorter in these models than 
in the F model. The fastest equilibration occurs in the KK 
model ($\beta=1/3$) where $z=3/2$. 

In calculations of the transient persistence probabilities, 
the initial configuration of the height variables
is taken to be perfectly flat, i.e. $h_{j}(t_{0})=0$ ($j=1,L$). The 
lattice size was in the range $10^4 \le L \le 10^6$, 
and the duration of the deposition process, measured in 
units of number of grown monolayers (ML), was $\sim 10^{3}$. 
The results were averaged over $\sim 10^{3}$ independent runs. 
For measurements in the steady--state situation, a saturation of the 
interface roughness was first obtained by depositing a large number
(of the order of $L^{z}$) of monolayers and subsequent time evolution from
one of the steady--state configurations obtained this way was used for 
measuring the persistence probabilities. A  much smaller lattice length 
($L=1000$ for the F model, $L=500$ for the KK model, $L=200$ for the KPK model, 
and $L=40$ for the LC, WV, DT and CKD models) was used in 
these calculations in order to reach the steady--state saturation 
within reasonable simulation times. 

\begin{table}
\begin{tabular}{|c|cc|cc|cc|cc|cc|}
\hline \hline Growth model & $L$ && $\theta_{+}^{T}$ && $\theta_{-}^{T}$ 
&& $\beta$ && Universality class & \\ \hline \hline 
F & $10^{6}$  && $1.57 \pm 0.10$ && $1.49 \pm 0.10$ && $0.25 \pm 0.01$ && EW &
\\ \hline KK & $5 \times 10^{4}$  && $1.68 \pm 0.02$ && $1.21 \pm
0.02$ && $0.33 \pm 0.01$ && KPZ &
\\ \hline LC & $10^{4}$  && $0.84 \pm 0.02$ && $0.84 \pm 0.02$ && $0.37 \pm 0.01$ && MH &
\\ \hline WV & $10^{4}$  && $0.94 \pm 0.02$ && $0.98 \pm 0.02$ && $0.37 \pm 0.01$ && MBE &
\\ \hline DT & $10^{4}$  && $0.95 \pm 0.02$ && $0.98 \pm 0.02$ && $0.38 \pm 0.01$ && MBE &
\\ \hline CKD & $10^{4}$  && $0.98 \pm 0.02$ && $0.93 \pm 0.02$ && $0.35 \pm 0.01$ && MBE &
\\ \hline KPK & $10^{4}$  && $1.04 \pm 0.02$ && $1.01 \pm 0.02$ && $0.31 \pm 0.01$ && MBE & 
\\ \hline \hline
\end{tabular}
\caption{Positive and negative persistence exponents, $\theta_{+}$ and 
$\theta_{-}$, for the transient ($T$) regime, measured for seven 
different discrete growth models (identified in the first column) 
using kinetic Monte Carlo simulations with relatively large system 
sizes ($L$). The measured growth exponent, $\beta$, and the 
universality class of the model are indicated in the last two 
columns, respectively.}   
\label{tab_TR}
\end{table}

\begin{figure}
\includegraphics[height=7.5cm,width=10cm]{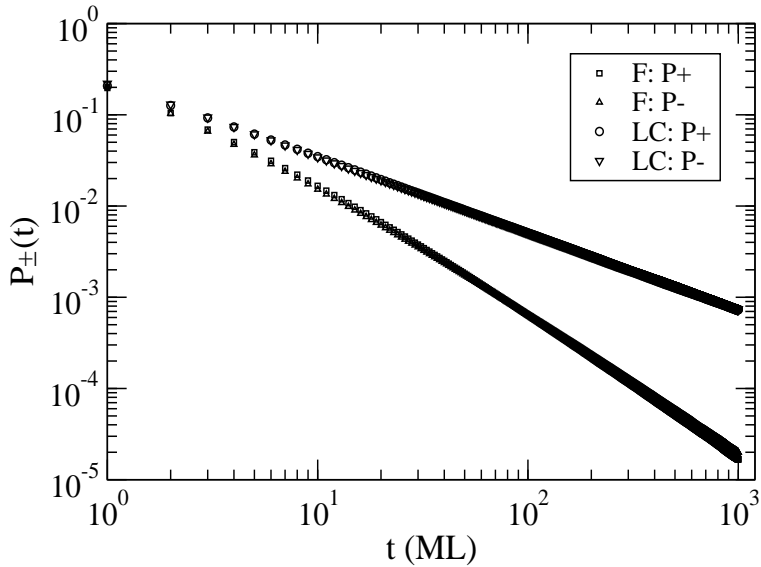}
%\vspace{0.5cm}
\caption{\label{fig1} 
Transient persistence probability for 
the (1+1)--dimensional linear F and LC growth models. As expected, 
the positive and negative persistence probabilities are identical in 
these models. The system size is $L=10^{6}$ for the F model and 
$L=10^{4}$ for the LC model, and an average over $10^{3}$ 
independent runs was performed. The slopes of the double--log
plots yield the values of the
transient persistence exponents shown in Table~\ref{tab_TR}.}
\end{figure}

\begin{figure}
\includegraphics[height=7.5cm,width=10cm]{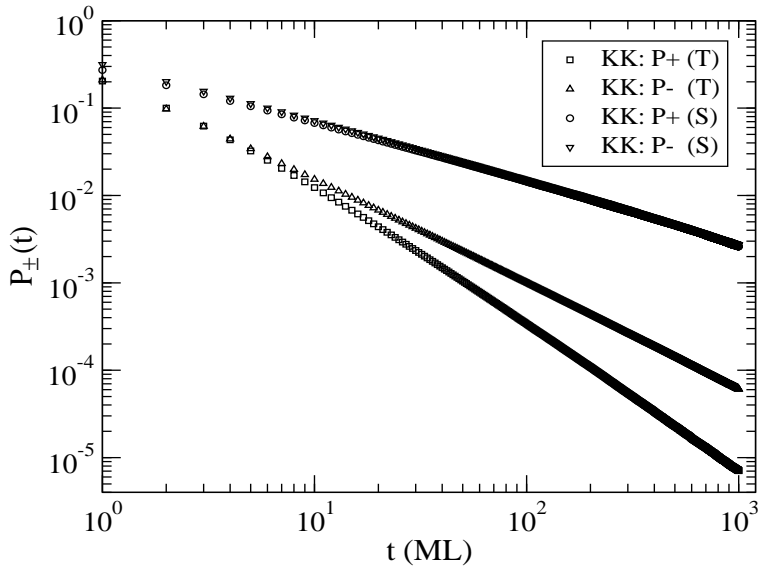}
\caption{\label{fig2} 
Positive and negative transient (bottom two curves) and steady--state 
(top two curves, mostly overlapped) persistence probabilities 
for the (1+1)--dimensional RSOS KK model. The faster 
decay of the positive persistence probability in the transient regime is 
due to the negative sign of $\lambda_{2}$ in the equivalent 
continuum equation of Eq.~(\ref{KPZ_eq}). In the 
transient case, systems of size $L=5 \times 10^{4}$ 
were averaged over $5 \times 10^{3}$ independent runs. The 
steady--state simulation was done for $L=500$ and a similar 
average was performed.}
\end{figure}

\begin{figure}
\includegraphics[height=7.5cm,width=10cm]{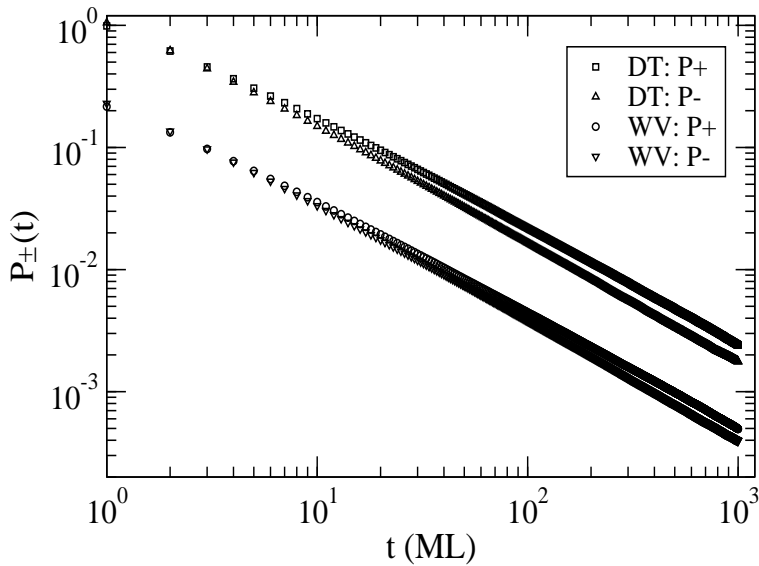}
\caption{\label{fig3} 
Positive and negative transient persistence probabilities 
for the (1+1)--dimensional nonlinear DT and WV growth models. 
We note that despite the difference in their local diffusion rules, these 
two models behave identically as far as the transient persistence 
probability is concerned. The curves corresponding to the DT model have 
been shifted upward in order to avoid a complete overlap of the plots for the 
two models. The system size is $L=10^{4}$ and an average over 
$10^{3}$ independent runs was performed. The slopes of the double--log plots 
yield the transient persistence exponents given in Table~\ref{tab_TR}.}
\vspace{1.0cm}
\end{figure}

\begin{figure}
\includegraphics[height=7.5cm,width=10cm]{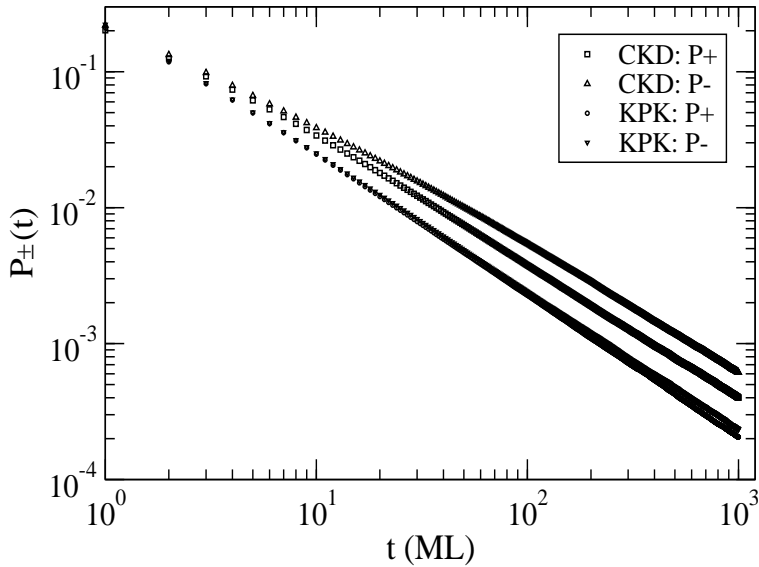}
\caption{\label{fig4}
Positive and negative transient persistence probabilities for the 
(1+1)--dimensional CKD (the upper two curves) and RSOS (the lower two, 
almost overlapped curves) models that belong to MBE universality class. 
In both cases the system size was $L=10^{4}$ and an average over
$10^3$ independent runs was performed. The slopes of the double--log plots
yield the transient persistence exponents given in Table \ref{tab_TR}.}
\end{figure}

The positive (negative) persistence probabilities in both 
transient and steady--state regimes were obtained as 
the fraction of sites that maintain the values of their heights 
persistently above (below) their initial values, averaged 
over a large number ($\sim 10^4$) of independent runs. 
The persistence exponents were obtained from power law 
fits to the decay of these probabilities, as shown in 
Figs.~\ref{fig1}--\ref{fig4} and \ref{fig6}--\ref{fig8} 
for the transient and steady--state regimes, respectively.

\begin{table}
\begin{tabular}{|c|cc|cc|cc|cc|cc|cc|}
\hline \hline Growth model & $L$ && $\theta_{+}^{T}$ && $\theta_{-}^{T}$ &&
$\theta_{+}^{S}$ && $\theta_{-}^{S}$ && $\beta$ &
\\ \hline \hline F & $10^{3}$  && $1.67 \pm 0.10$ && 
$1.56 \pm 0.10$ && $0.78 \pm 0.02$ && $0.76 \pm 0.02$ && $0.25 \pm 0.01$ &
\\ \hline KK & $5 \times 10^{2}$  && $1.70 \pm 0.02$ && $1.27 \pm 0.02$ && 
$0.71 \pm 0.02$ && $0.71 \pm 0.02$ && $0.30 \pm 0.01$ &
\\ \hline LC & $40$  && $0.98 \pm 0.02$ && $0.96 \pm 0.02$ && 
$0.67 \pm 0.02$ && $0.67 \pm 0.02$ && $0.32 \pm 0.01$ &
\\ \hline WV & $40$  && $0.94 \pm 0.02$ && $0.99 \pm 0.02$ &&
$0.65 \pm 0.02$ && $0.70 \pm 0.02$ && $0.35 \pm 0.01$ &
\\ \hline DT & $40$  && $0.98 \pm 0.02$ && $1.01 \pm 0.02$ &&
$0.64 \pm 0.02$ && $0.72 \pm 0.02$ && $0.36 \pm 0.01$ &
\\ \hline CKD & $40$  && $1.11 \pm 0.02$ && $0.99 \pm 0.02$ &&
$0.78 \pm 0.02$ && $0.66 \pm 0.02$ && $0.33 \pm 0.01$ &
\\ \hline KPK & $2 \times 10^{2}$  && $1.16 \pm 0.02$ && $1.09 \pm 0.02$ &&
$0.70 \pm 0.02$ && $0.68 \pm 0.02$ && $0.28 \pm 0.01$ & \\ \hline \hline
\end{tabular}
\caption{\label{tab_short} Positive and negative persistence exponents, 
$\theta_{+}$ and $\theta_{-}$, for the transient ($T$) and the steady state 
$(S)$ regimes of our seven different discrete growth 
models, obtained from simulations with relatively small samples sizes ($L$).
To illustrate the effects of 
reduced system sizes on the measured exponents, we have shown the values of 
$\beta$ obtained from these simulations in the last column.}
%\vspace{1.5cm}
\end{table}

For all the models studied here, we have also measured the value of
the growth exponent $\beta$ in both transient and steady--state
simulations. Since the latter simulations were carried out for smaller
values of the system size $L$, these measurements provide useful
information about the dependence of the measured exponent values 
on the lattice size. 
Similar information is also provided by the values of the
transient persistence exponents obtained from measurements 
in the initial stage of the steady--state simulations.
The transient exponent values obtained from the large--$L$ 
simulations are listed in Table~\ref{tab_TR}, and both transient and 
steady--state exponent values obtained from simulations of relatively
small samples are shown in Table~\ref{tab_short}. The measured values
of the growth exponent $\beta$ are also shown in these Tables.  

Estimation of the probable error in the measured values of the 
growth and persistence exponents is a delicate task (and surely
depends on precisely how the exponent error is defined), since there is
not a traditional accepted method to evaluate the error in dynamical 
simulations. To solve this problem we did the following simulations. 
We decreased the number of independent runs used for the averaging 
procedure by a factor of $2$, keeping the size of the system
constant. Under these circumstances, we have measured the exponents 
corresponding to the two different numbers of independent runs 
and the differences between the obtained values of the exponents 
were used as error estimates for $\beta$ and $\theta$, 
respectively. Approximately the same size of the error bar 
was obtained from the estimations of fluctuations 
in the value of the local slope of the double--log plots. 
We have also noticed that a reduction of the lattice 
size (imposed for the steady--state persistence calculations) 
produces lower values of the growth exponents, as shown in Table 
\ref{tab_short}. This is because the downward bending (approach to
saturation) of double--log width versus time plots occurs at 
shorter times in simulations of smaller systems. However, the 
smaller--$L$ simulations seem to lower the measured values of 
the growth exponents by a maximum of about 10\% percent. So we 
conclude that this effect is not dramatic and that the steady--state 
results reported below are reliable.   

The measured values of $\beta$ agree reasonably well with the 
expected ones (see Section \ref{A}) within their errors. As expected,
the agreement is better in the case of larger values of $L$.
For the larger--$L$ simulations ($L \sim 10^{4}$), we have 
found that the growth exponents of the  F, LC and KK models are in 
excellent agreement with their corresponding expected values of 
$1/4$, $3/8$ and $1/3$, respectively (see 
Table~\ref{tab_TR}). The DT and WV models are found to behave 
similarly at early (transient) stages of their interface growth, 
at least in (1+1)--dimensions, their growth exponents being: 
$\beta_{WV} \approx 0.37$ and $\beta_{DT} \approx 0.38$. 
The closeness of these values to the value of $3/8$, which 
corresponds to the MH universality class, suggests that the 
nonlinear term that appears in the associated dynamic equation 
(i.e. Eq.~(\ref{MBE_eq})) has a very weak effect for the range 
of lattice sizes used in our study. In addition, we have found 
that the CKD model characterized by the nonlinear coefficient 
$\lambda_{22}=2$ and control parameter $c=0.02$ has a growth 
exponent $\beta_{CKD} \sim 0.35$, in agreement with 
Ref.~\cite{chandan}. These particular parameter values 
ensured the elimination of any interfacial instability, 
thus allowing a calculation of the steady--state persistence 
properties. Regarding the conserved KPK model, we have 
observed that the growth exponent has a value that is 
slightly smaller than $1/3$, a result that agrees with 
Ref.~\cite{KPK}. 

The temporal behavior of the transient persistence probability 
in our models is shown in Figs.~\ref{fig1}--\ref{fig4}. From these 
measurements, we obtained the transient persistence exponents 
by fitting the linear middle regions (excluding the small--$t$ and large--$t$ 
ends, typically using the data for $20<t<800$) of the double--log plots to 
straight lines. As expected, due to the invariance 
of the interfaces of the F and LC models (which are 
characterized by $linear$ continuum equations) under a 
change of sign of the height variables, we obtained equal 
positive and negative transient persistence exponents within 
the error bars, as displayed in Fig.~\ref{fig1}. However, 
we mention that the F model has a rather slow convergence 
of the positive and negative exponents towards their 
long--time value of $\sim 1.55$ observed in much longer 
simulations. The results for F and LC models, that correspond 
to $\beta=1/4$ and $3/8$, respectively, agree well with the values 
reported by Krug {\it et al.} \cite{Krug1}. The same level of 
agreement is also found in the case of the KK model 
\cite{Krug2}, shown in Fig.~\ref{fig2}, for which the transient
persistence exponents are $\theta_{+}^{T} \approx 1.68$ and 
$\theta_{-}^{T} \approx 1.21$ in (1+1)--dimensions. We note  
that the negative persistence probability has a slower decay 
than the positive one. This is due to the constant coefficient, 
$\lambda_{2}$, of the nonlinear term 
$|{\bf \nabla} h({\bf r},t)|^{2}$ of the KPZ equation 
(which provides a continuum description of the KK model) 
having a negative sign \cite{Krug2}. 

\begin{figure}
\includegraphics[height=15cm,width=11cm]{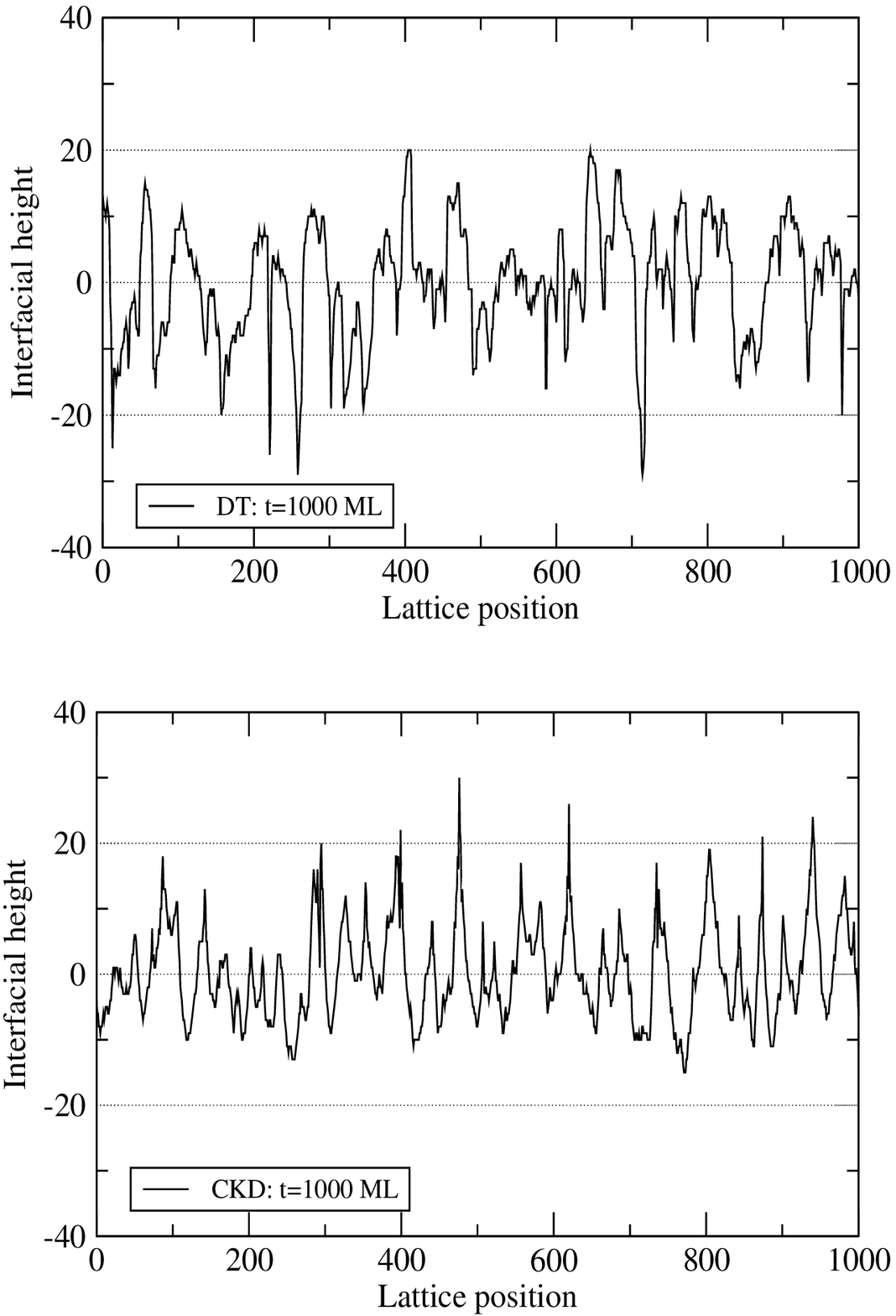}
\caption{\label{fig5} 
Morphologies of the (1+1)--dimensional DT (top) and CKD (bottom) 
stochastic models for $L=10^{4}$ (only a  portion of 1000 sites is shown) 
and $t=10^{3}$ ML. In the DT model, we notice a breaking of up--down 
symmetry due to the formation of deep grooves, while in the CKD model, the 
representative asymmetric feature corresponds to high pillars.} 
\end{figure}

\begin{figure}
%\vspace{1.0cm}
\includegraphics[height=7.5cm,width=11cm]{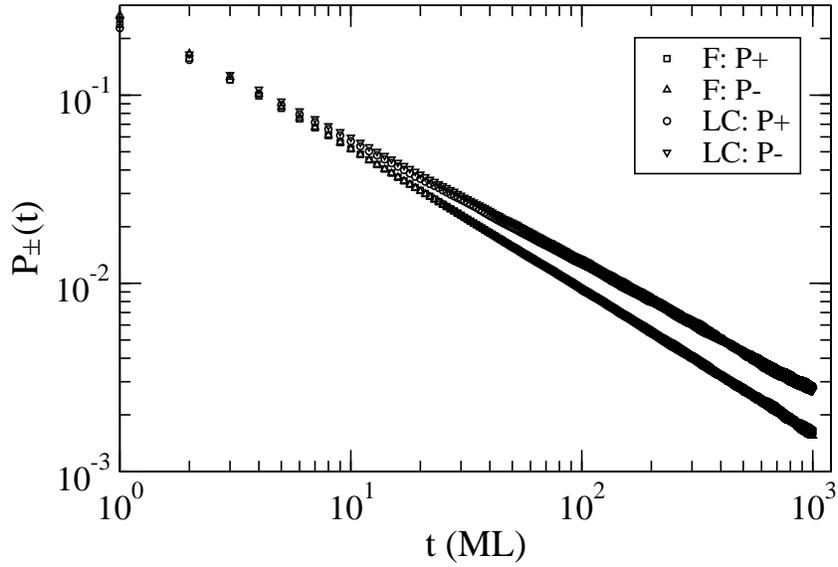}
\caption{\label{fig6} 
Positive and negative steady--state 
persistence probabilities for the (1+1)--dimensional F and LC models which are 
governed by linear continuum dynamical equation. 
The temporal decay of the persistence probability 
is slower in the LC model which has a larger growth exponent 
($\beta_{LC} = 3/8$, $\beta_{F} = 1/4$). We used 
$L=1000$ and $t_{0}=4 \times 10^{6}$ ML for the F model, 
and $L=40$, $t_{0}=10^{6}$ ML for the LC model. 
The displayed results were averaged over 5000 
independent runs. The measured slopes of the double--log plots yield the 
steady--state persistence exponents shown in Table \ref{tab_short}.} 
\vspace{1.0cm}
\end{figure}

\begin{figure}
\includegraphics[height=7.5cm,width=10cm]{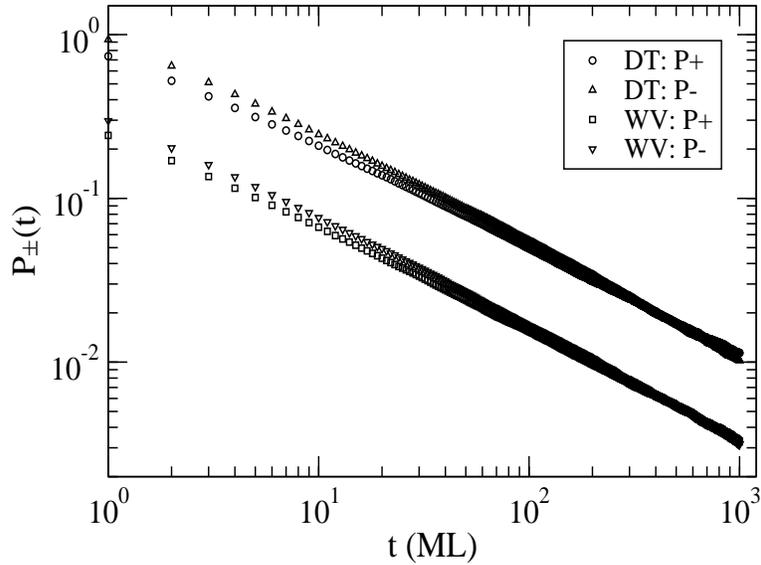}
\caption{\label{fig7} 
Steady--state persistence probabilities for 
two (1+1)--dimensional models in the MBE universality class -- 
the DT and WV models. As in the 
transient case, these two models exhibit almost identical 
persistence behavior in the steady state. The effects of the
nonlinearity in their continuum dynamical description are not very 
prominent for the small lattice sizes considered here. 
For the data shown, systems of size $L=40$ were 
equilibrated for $t_{0}=10^{5}$ ML, and the results were 
averaged over 5000 independent runs. The persistence 
plots for the DT model have been shifted up in order  
to make them distinguishable from the WV plots.
The measured slopes of the double--log plots yield the
steady--state persistence exponents shown in Table \ref{tab_short}.} 
\end{figure}

\begin{figure}
\includegraphics[height=7.5cm,width=10cm]{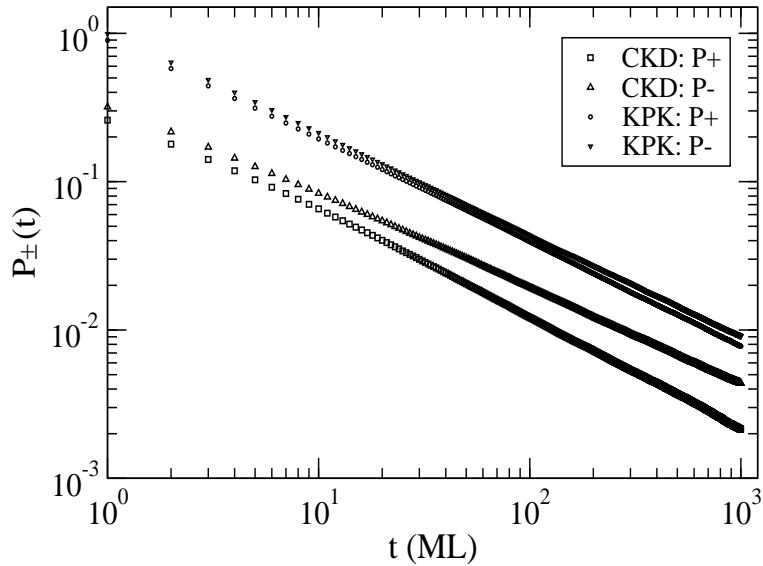} 
\caption{\label{fig8} Double--log plots of the steady--state 
persistence probabilities of (1+1)--dimensional MBE class 
CKD and KPK (shifted up by a constant amount) models. 
While the KPK model does not show a clear effect of 
nonlinearity in the values of the persistence exponents, the CKD 
model shows  positive and negative persistence 
exponents that are clearly different from each other. 
Systems of size $L=40$ (CKD) and $L=200$ (KPK) were 
equilibrated for $t_{0} \sim 10^{5}$ ML. The results were 
averaged over $10^{4}$ independent runs. 
The measured slopes of the double--log plots yield the
steady--state persistence exponents shown in Table \ref{tab_short}.}
%\vspace{1.0cm}
\end{figure}

For the models described by the fourth--order nonlinear MBE 
equation (i.e. WV, DT, CKD and KPK models), we expect to find different 
positive and negative transient persistence exponents due to the 
fact that their morphologies violate the up--down interfacial 
symmetry with respect to the average level. No information 
about how different these two exponents should be is available in the
literature. In most of these growth models, we 
observe that the two exponents are not very different from each 
other, especially during the transient regime. Fig.~\ref{fig3} 
shows the transient regime results for DT and WV models, which 
are indeed very similar -- their persistence probability curves 
have almost identical behavior. We note here that the negative 
persistence probability has a faster decay than the positive 
persistence probability. This indicates a negative sign of  
$\lambda_{22}$, the coefficient that multiplies the nonlinear term 
$\nabla^{2}|{\bf \nabla} h({\bf r},t)|^{2}$ of the MBE equation. 
However, the relative order of the values of these exponents is 
reversed when $\lambda_{22}>0$, which is the case in the CKD and KPK 
models, as shown in Fig.~\ref{fig4}. To clarify this aspect, we show in 
Fig.~\ref{fig5} the interfacial morphologies of DT and CKD 
models. We used a lattice of $L=10^{4}$ sites (but only a portion of 
1000 sites is shown in each case) and the displayed configurations 
correspond to a time of $10^{3}$ ML. The interface of the DT model is 
characterized by deep grooves, while the profile in the CKD model exhibits 
the distinct feature of high pillars. Both morphologies display strong 
up--down interfacial asymmetry, but their representative 
features (i.e. deep grooves and high pillars) are opposite in ``sign'',
indicating a reversal of the sign of the coefficient $\lambda_{22}$ (note
that a reversal of the sign of $\lambda_{22}$ in Eq.~(\ref{MBE_eq}) 
is equivalent to changing the sign of the height variable 
$h({\bf r}, t)$).

As summarized in Table~\ref{tab_TR}, the DT, WV and CKD models show
very similar values for the transient persistence exponents when the 
above mentioned effect of the sign of $\lambda_{22}$ is taken into 
account. However, some deviation from the exponent values for this 
group of models is observed in the RSOS KPK model which shows the 
smallest difference between the positive and negative persistence 
exponents. Finite size effects appear to be stronger in this case.
These effects also cause an increase in the measured values of 
the persistence exponents above the expected values. A similar 
behavior is found in the steady--state results as well, as
described below. 

Our calculations of WV, DT, CKD and KPK persistence exponents 
illustrate the feasibility of studying 
this type of nonequilibrium statistical probabilities 
for a large class of nonequilibrium applications described 
by nonlinear dynamical equations. Until now, the only nonlinear 
equation for which  persistence exponents have been 
calculated \cite{Krug2} is the KPZ equation which is 
arguably the simplest nonlinear Langevin equation. Further, 
the nonlinearity in the KPZ equation becomes  irrelevant 
in the steady--state regime in (1+1)--dimensions. So, the effects 
of nonlinearity are not reflected in the steady--state persistence
behavior of (1+1)--dimensional KPZ systems. An immediate
concern would be that more complex nonlinear dynamic 
equations might be less approachable from the point of 
view of persistence probability calculation. Our results 
for four nonlinear models eliminate this possibility 
and illustrate the applicability and usefulness of persistence 
probability calculations in the study of surface fluctuations.

Figures \ref{fig6}--\ref{fig8} display our results for the 
steady--state persistence probabilities. The values of the 
growth and persistence exponents obtained from the steady--state 
simulations are summarized in Table~\ref{tab_short}. The values 
of the steady--state persistence exponents in the F and LC 
models, corresponding, respectively,  to the $\nabla^{2}$ 
and $\nabla^{4}$ linear equations, (see Fig.~\ref{fig6}) 
are consistent with the values of the corresponding growth 
exponents (as predicted by Eq.~(\ref{theta_S})) obtained 
from the same small--$L$ simulations. For the WV and DT 
models, as shown in Fig.~\ref{fig7}, we obtain very similar 
positive and negative persistence exponents. In the case of 
the KK model we find, as expected, identical positive and 
negative exponents ($\theta_{\pm}^{S} \approx 0.71$),  
as shown in Fig.~\ref{fig2}.

Among the models belonging to the MBE universality class, the
KPK model exhibits steady--state persistence exponents that are
systematically higher than the ones obtained for the remaining three 
(WV, DT and CKD) models. Our study of the dynamical scaling 
behavior of the KPK model indicates that both $\alpha$ 
($\sim 0.9$) and $z$ ($\sim 2.9$) in this model are reasonably close to 
the expected values, in agreement with Ref.~\cite{KPK}. 
Therefore, the reason for the differences between the values of the 
steady--state persistence exponents for the KPK model and those in the
other models in the MBE universality class is unclear. This
discrepancy may very well be arising from subtle differences in finite
size (and time) effects in the simulations for persistence exponents
and dynamic scaling. Further investigation of the applicability of 
this RSOS model in understanding MBE growth is beyond the scope of 
the present study.

As shown in Fig.~\ref{fig8}, a nice illustration of the presence of
nonlinearity in the underlying dynamical equation is 
provided by the steady--state persistence exponents of the 
CKD model, characterized by distinct values, 
$\theta_{+}^{S} \approx 0.78$ and 
$\theta_{-}^{S} \approx 0.66$, of the positive and negative
exponents. Although one must take into account the fact that 
these values might be slightly overestimated (by approximatively 
5\%) due to the smallness of the sample sizes used in the
steady--state simulations, these exponents provide a good 
qualitative account of the nontrivial up--down asymmetric 
persistence behavior expected for nonlinear models belonging 
to the MBE universality class.

\begin{table}
\begin{tabular}{|c|cc|cc|cc|}
\hline \hline $~~L~~$ & $\theta_{+}^{T}$ && $\theta_{-}^{T}$ && $\beta$ & 
\\ \hline \hline $10^{4}$ & $0.95 \pm 0.02$ && $0.98 \pm 0.02$ && $0.38 \pm 0.01$  &
\\ \hline $10^{2}$  & $0.96 \pm 0.02$ && $0.99 \pm 0.02$ &&  $0.37 \pm 0.01$ &
\\ \hline $40$  & $0.98 \pm 0.02$ && $1.01 \pm 0.02$ && $0.36 \pm 0.01$ & 
\\ \hline \hline
\end{tabular}
\caption{\label{tab_DT} Transient positive and negative 
persistence exponents, $\theta_{\pm}^{T}$, obtained for the 
DT model with different system sizes ($L$). The effect of the 
system size on the measured growth exponent, $\beta$, is 
displayed in the last column. No result for steady--state persistence 
exponents is available for system sizes larger than $\sim 100$, 
due to the impossibility of reaching saturation of the 
interface width for such values of $L$ in time scales accessible in
simulations. The results shown here
were averaged over 500 (for $L=10^{4}$), $5\times 10^{4}$ (for 
$L=100$) and $10^{5}$ (for $L=40$) independent runs.}
\end{table}

Next we investigate the influence of small sample sizes 
on the measured values of the persistence exponents. 
We mainly used the DT model to answer 
this question and we pursued the following two tests. First we 
decreased the size of the system from $10^{4}$ to $100$ and then to $40$ 
and found the values of the growth and persistence exponents, as 
summarized in Table~\ref{tab_DT}. We note that as the lattice 
length decreases to $40$, the persistence exponents increase by 
$\sim$ 2\%, while the growth exponents increase by $\sim$ 5\%.
As a second test, we have applied the noise reduction 
technique to both the DT and WV models. It has been shown \cite{nrt3} that a noise 
reduction factor of $m=5$ helps the DT model to recover quite accurately 
the universal exponents corresponding to the MBE universality 
class. In addition, the noise reduced WV model exhibits, at late 
evolution times, its true EW asymptotic universality, which is
difficult to observe without applying noise reduction. Therefore, 
the DT model with the appropriate noise reduction factor 
is expected to provide the correct persistence exponents 
associated with the fourth--order nonlinear dynamical equation for
MBE growth. The results obtained from the simulations with 
noise reduction are summarized in Table~\ref{tab4}. We notice 
that the noise reduction scheme produces only a minor change 
in the persistence exponents and in addition, the results 
obtained for $m=5$ agree within the error bars with those 
for the CKD model. We, therefore, conclude that the noise 
reduced DT model and the discrete CKD model provide a good 
representation of the MBE universality class, characterized by two 
different steady--state persistence exponents: 
$\theta_{+}^{S} \sim 0.66$ (positive persistence) and 
$\theta_{-}^{S} \sim 0.78$ (negative persistence). These 
nontrivial persistence exponents for this class have not been 
reported earlier, and it would be useful to check these results 
from further theoretical or experimental studies. Regarding the noise
reduced WV model we mention that the convergence of $\theta^S$ towards
the expected value of $3/4$ is rather slow in the case of the positive
exponent and probably a higher value of the noise reduction factor
would be necessary to reveal the true EW universality. We did not
explore this technical issue any further.

\begin{table}
\begin{tabular}{|c|cc|cc|cc|}
\hline \hline Growth model & $m$ && $\theta_{+}^{S}$ && $\theta_{-}^{S}$ &
\\ \hline \hline DT  & $1$ && $0.64 \pm 0.02$ && $0.72 \pm 0.01$  &
\\ \hline DT  & $5$ && $0.65 \pm 0.02$ &&  $0.77 \pm 0.01$ &
\\ \hline WV  & $1$ && $0.65 \pm 0.02$ &&  $0.70 \pm 0.01$ &
\\ \hline WV  & $5$ && $0.68 \pm 0.02$ && $0.75 \pm 0.01$ & 
\\ \hline \hline
\end{tabular}
%\vspace{0.5 cm} 
\caption{\label{tab4} Positive and negative persistence 
exponents, $\theta_{\pm}^{S}$, for the steady state of
the DT and WV models for two 
different values of the noise reduction factor, $m$. Systems of 
size $L=40$ were equilibrated for $10^{5}$ ML and the results 
were averaged over $5000$ independent runs.}
\end{table}

We note that among the positive 
and negative steady--state persistence exponents for these
nonlinear growth models, the smaller one (for 
example, the positive exponent in the DT model or the negative 
exponent of the CKD model), turns out to be close to $(1-\beta)$. 
In the next subsection, we show analytically 
that this relation between the smaller steady-state persistence exponent
and the dynamic growth exponent is, in fact, exact.
Our numerical studies suggest a connection of this result with
the morphology that develops in the steady--state regime. 
As shown in Fig.~\ref{fig5}, the characteristic feature of the DT 
morphology is the presence of deep grooves, while the CKD model 
exhibits high pillars. Loosely speaking, in the case of 
the DT model, we expect the relation of Eq.~(\ref{theta_S}) 
to be more likely to be satisfied by the positive persistence 
exponent than the negative one because the preponderant grooves, 
responsible for the negative persistence exponent, represent the 
effects of the nonlinearity of the underlying MBE dynamics. More
work is clearly needed for a better understanding of the possible
relationship between such ``nonlinear'' features of the interface 
morphology and the value of the persistence exponent.

\subsection{An exact relation between steady-state persistence 
exponents and the growth exponent}
\label{analytic}

As mentioned earlier, for interface heights $h({\bf r} ,t)$ evolving via 
a Langevin 
equation that preserves $(h\to -h)$ symmetry (for example, any linear Langevin 
equation), the steady state persistence exponents satisfy the scaling relation
$\theta_{+}^S=\theta_{-}^S= 1-\beta$, where $\beta$ is the growth 
exponent~\cite{Krug1}. 
In this subsection, we derive a generalized scaling relation,
\begin{equation}
\beta= {\rm max}\left[1-\theta_{+}^S, 1-\theta_{-}^S\right],
\label{sc1}
\end{equation}
which is valid even in the absence of $(h\to -h)$ symmetry. When 
this
symmetry is restored, Eq.~(\ref{sc1}) reduces to the known result~\cite{Krug1},  
$\theta_{+}^S=\theta_{-}^{S}=1-\beta$.

To derive the relation in Eq.~(\ref{sc1}), we start with a generic interface 
described by a height field $h({\bf r},t)$ and define the relative height,
$u({\bf r},t)=h({\bf r},t)-{\overline h({\bf r},t)}$ where ${\overline 
h({\bf r},t)}=\int h({\bf r},t)d{\bf r}/V$
is the spatially averaged height and $V$ is the volume of the sample. 
Let us also define the incremental
auto-correlation function in the stationary state,
\begin{equation}
C(t,t')= \lim_{t_0\to \infty}\langle [u({\bf r},t+t_0)-u({\bf r},t'+t_0)]^2 \rangle.
\label{incre1}
\end{equation}
It turns out that for generic self-affine interfaces 
(which do not have to be Gaussian),
this function $C(t,t')$ depends only on the time 
difference $|t-t'|$ (and not on the 
individual times $t$ and $t'$) in a power-law 
fashion for large $|t-t'|$~\cite{Krug3,Krug2},
\begin{equation}
C(t,t')\sim |t-t'|^{2\beta},
\label{incre2}
\end{equation}
where $\beta$ is the growth exponent. 

This particular behavior of the auto-correlation function 
in Eq.~(\ref{incre2}) is typical of a fractional Brownian motion (fBm). 
A stochastic process $x(t)$ with zero mean
is called an fBm if its incremental correlation function
$C(t_1,t_2)=\langle [x(t_1)-x(t_2)]^2\rangle$ depends only on 
the time difference
$|t_1-t_2|$ in a power-law fashion for large arguments~\cite{MV},
\begin{equation}
C(t_1,t_2)=\langle [x(t_1)-x(t_2)]^2\rangle \sim |t_1-t_2|^{2H},
\label{fbm1}
\end{equation}
where $0<H<1$ is called the Hurst exponent of the fBm. 
For example, an ordinary Brownian motion
which evolves as $dx/dt=\eta(t)$ where $\eta(t)$ is a 
Gaussian white noise with zero mean and
a delta function correlator, satisfies Eq.~(\ref{fbm1}) with $H=1/2$. 
Thus an ordinary Brownian 
motion is a
fBm with $H=1/2$. It follows clearly by comparing 
Eqs.~(\ref{incre2}) and (\ref{fbm1}) that
the relative height $u({\bf r}, t)$ of a generic interface at a 
fixed point ${\bf r}$ in space, 
in its stationary state, is also a fBm with Hurst 
exponent, $H=\beta$. Note that an fBm is not necessarily Gaussian.

We are then interested in the `no return probability' to the 
initial value of the fBm process
$u({\bf r},t)$. So, the relevant random process 
is $Y({\bf r},t)= u({\bf r},t+t_0)-u({\bf r},t_0)$. 
Clearly, $Y({\bf r},t)$ 
is also
a fBm with the same Hurst exponent $\beta$ since the 
incremental correlation function of $Y$
is the same as that of $u({\bf r},t)$. We are then interested 
in the zero crossing
properties of the fBm $Y({\bf r},t)$. Now, consider the
process $Y({\bf r},t)$ as a function of time, at a fixed point 
${\bf r}$ in space, from time 
$t_0$
to time $t_0+t$ where $t_0\to \infty$. There are two types of intervals
between successive zero crossings in time, 
the `$+$' type (where the process lies above
$0$) and the `$-$' type (where the process lies below $0$).

In general, the statistics of the two types of intervals 
are different. Only, in special cases,
where one has the additional knowledge that 
the process $Y({\bf r},t)$ is symmetric around $0$
(i.e., processes which preserve the $(h\to -h)$ symmetry), 
the
$+$ and $-$ intervals will have the same statistics. 
For such cases, a simple scaling argument
was given in Ref.~\cite{Krug1} to show that the length of an 
interval of either type has a
power-law distribution, $Q(\tau)\sim \tau^{-1-\theta^S}$ 
(for large $\tau$) with
$\theta^S=1-H=1-\beta$. Note that this relation between the 
persistence exponent
and the Hurst exponent is very general and holds for any symmetric fBm, i.e., 
any 
stochastic process with zero mean
(not necessarily Gaussian) satisfying Eq.~(\ref{fbm1}). 
Recently, other applications 
of this result have been found~\cite{SM1,Krug4}.
For general nonsymmetric processes, however, one would expect that
$Q_{\pm}(\tau)\sim \tau^{-1-\theta_{\pm}^S}$ for large
$\tau$,
where $\theta_{+}^S$ and $\theta_{-}^S$ are, in general, different. 
Here we generalize this
scaling argument of Ref.~\cite{Krug1} (derived for a symmetric process) 
to include the 
nonsymmetric cases and derive the result in Eq.~(\ref{sc1}).

The derivation of Eq.~(\ref{sc1}) follows more or less the same 
line of arguments as that used
in Ref.~\cite{Krug1} for the symmetric case. 
Let $P(Y,\tau)$ denote the probability
that the process has value $Y$ at time $\tau$, given that 
it starts from its initial value $0$ at
$\tau=0$. Then, it is natural to
assume that the normalized probability distribution 
$P(Y,\tau)$ has a scaling form,
\begin{equation}
P(Y,\tau)= {1\over {\sigma (\tau)}}f\left({Y\over {\sigma(\tau)}}\right),
\label{sf1}
\end{equation}
where $\sigma(\tau)$ is the typical width of the process, 
$\sigma^2(\tau)= \langle 
Y^2(\tau)\rangle$.
It follows from Eq.~(\ref{incre2})
that $\sigma(\tau)\sim \tau^{\beta}$ for large $\tau$. 
The scaling function $f(z)$ is a
constant at $z=0$, $f(0)\sim O(1)$ (note that, in general, 
$f(z)$ is {\it not} a 
symmetric function of $z$) and should decrease
to $0$ as $z\to \pm \infty$. So, given that a zero occurs 
initially, the probability $\rho(\tau)=P(0,\tau)$ that the process
will return to $0$ after time $\tau$ (not necessarily for the 
first time) scales as
\begin{equation}
\rho(\tau)\sim {1\over {\sigma (\tau)}}\sim \tau^{-\beta},
\label{rho1}
\end{equation}
as $\tau\to \infty$. This function $\rho(\tau)$ indeed is the 
density of zero crossings between $\tau$ and $\tau+d\tau$. 
Thus, the total number of zeros upto a time $T$ is simply the integral,
\begin{equation}
N(T)= \int_0^{T} \rho(\tau)d\tau \sim T^{1-\beta},
\label{n1}
\end{equation}
for large $T$.

Next, we relate the persistence probabilities to the number of zeros.
Let $P_{\pm}(\tau)$ denote the probabilities that the process stays
positive (or negative) over the interval $[0,\tau]$, given that
it started from a zero. By definition, we have $P_{\pm}(\tau)\sim \tau^{-\theta_{\pm}^S}$
for large $\tau$. Then, $Q_{\pm}(\tau)=-dP_{\pm}(\tau)/d\tau\sim \tau^{-1-\theta_{\pm}^S}$
(as $\tau\to \infty$) denote the
probabilities that the process will cross zero next time 
(from the positive or the negative side respectively) between 
time $\tau$ and $\tau+d\tau$. Thus, $Q_{\pm}(\tau)$ are also 
the distribution of intervals of the two types of length $\tau$.

Now, consider a total length of time $T$. Let $N(T)$ denote the total number of intervals
in this period, half of them are $+$ types and the other half $-$ types, $N_{\pm}(T)=N(T)/2$.
Let, $n_{\pm}(\tau)$ denote the number of $\pm$ intervals of length $\tau$ within the period
$T$. Thus, the fraction of $+$ (or $-$) intervals of length $\tau$,
$n_{\pm}(\tau)/N_{\pm}(\tau)$,
by definition, are the two distributions $Q_{\pm}(\tau)$ 
provided $T$ is large. Thus, for large $T$, we have
\begin{equation}
n_{\pm}(\tau,T)= {{N(T)}\over {2}} Q_{\pm}(\tau)\sim N(T) \tau^{-1-\theta_{\pm}^S},
\label{sc2}
\end{equation}
for $1<<\tau \le T$. On the other hand, we have the length 
conservation condition (the total length covered by the intervals
must be $T$),
\begin{equation}
\int_0^{T} d\tau\, \tau \left[n_{+}(\tau) + n_{-}(\tau)\right]=T.
\label{sc3}
\end{equation}
Substituting the asymptotic behavior of $n_{\pm}(\tau)$ in 
Eq.(\ref{sc2}) into the left-hand side
of Eq.(\ref{sc3}), we get
\begin{equation}
N(T)\left[ { {T^{1-\theta_{+}^S}}\over {1-\theta_{+}^S}}+ { {T^{1-\theta_{-}^S}}\over
{1-\theta_{-}^S}}\right] \propto T.
\label{sc4}
\end{equation}
We next use $N(T)\sim T^{1-\beta}$ for large $T$ from Eq.(\ref{n1}). 
This gives, for large 
$T$,
\begin{equation}
\left[ { {T^{1-\theta_{+}^S}}\over {1-\theta_{+}^S}}+ { {T^{1-\theta_{-}^S}}\over
{1-\theta_{-}^S}}\right] \sim T^{\beta}.
\label{sc5}
\end{equation}
Taking $T\to \infty$ limit and matching the leading power of $T$ in 
Eq.~(\ref{sc5}), we arrive at our main result in Eq.~(\ref{sc1}).
Note that in the above derivation we have implicitly 
assumed a small-$t$  cut-off
and focused only on the distribution of large intervals.
Our numerical results obtained for a 
class of nonlinear interfaces in both $(1+1)$ and $(2+1)$ 
dimensions (see Sec. \ref{sim2} below) are consistent
with the analytical result in Eq.~(\ref{sc1}).

\subsection{Dependence of persistence probabilities on the initial 
configuration}
\label{sim-steady}

We present in this section some surprising simulation 
results  about the dependence of the persistence behavior 
(specifically, the values of the persistence exponents) on 
the choice of the initial configuration. In particular, we 
show that the steady--state exponents may be obtained
with a fair degree of accuracy from simulations in which 
the interface {\it has not} yet reached the steady state. We also 
present some results that have bearing on the measurability 
of the transient persistence exponents from experimental data. 

\begin{figure}
\includegraphics[height=12cm,width=16cm]{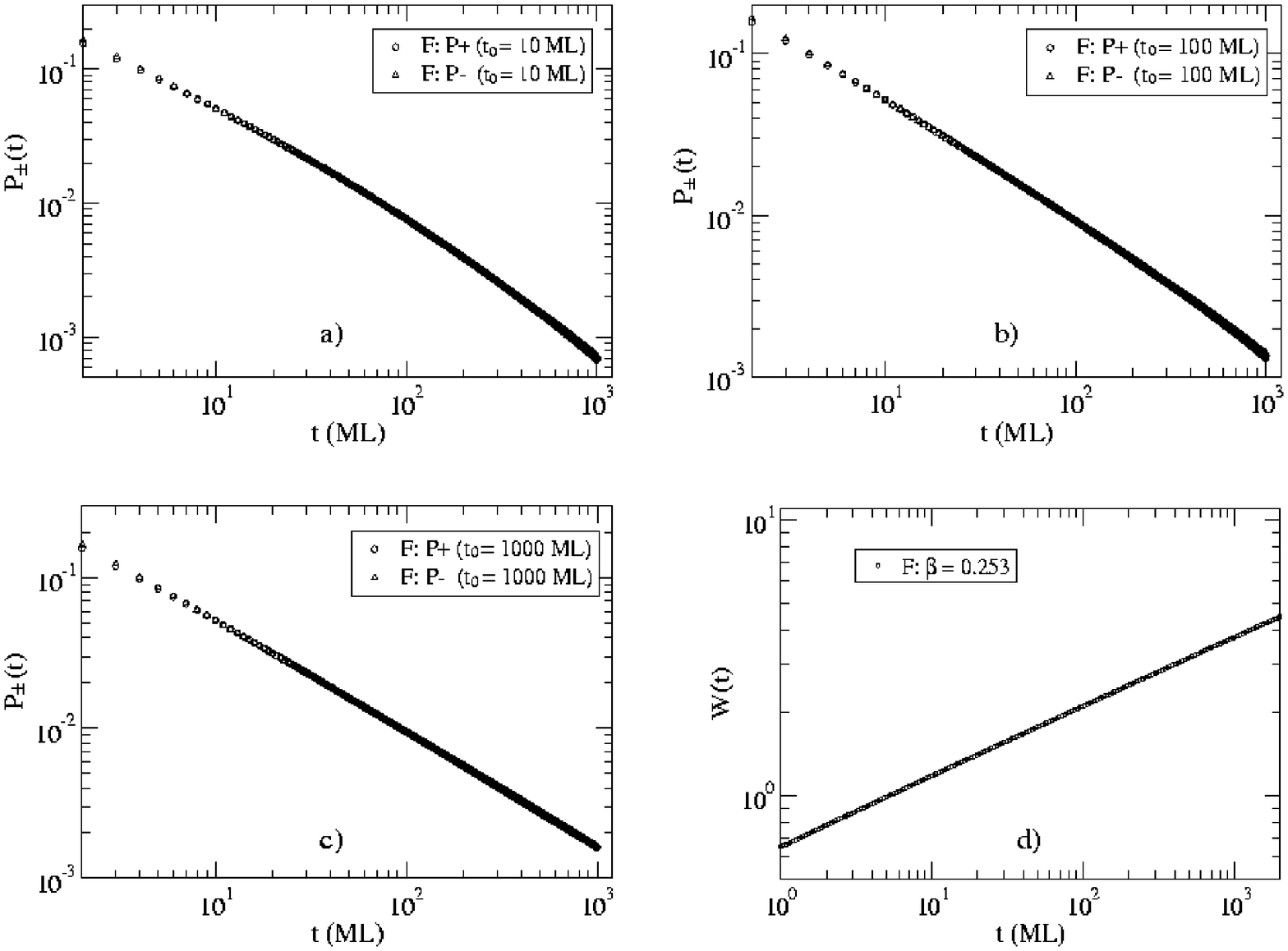}
\caption{\label{fig9} 
Log--log plots of the positive and negative persistence
probabilities [panels (a)--(c)] for the F model, obtained using different 
values of the initial time $t_0$. Systems with $L=10^{4}$ sites have 
been averaged over 500 independent runs. Persistence 
probabilities are computed starting 
from the configuration corresponding to: a) $t_{0}=10$ 
ML. We do not find a clear power law decay of the persistence 
curves. b) $t_{0}=100$ ML. As $t_{0}$ increases, a clearer power law 
behavior is observed. c) $t_{0}=1000$ ML. The power law 
decays are recovered and characterized by exponents in agreement 
with those corresponding to the steady--state regime: $\theta^{S}_{\pm} 
\approx 0.75$. d) Log--log plot of the interface width $W$ as a function
of time $t$ (in units of ML). The value of the slope 
(equal to the growth exponent $\beta$) agrees with the expected value, 
$\beta=0.25$.} 
\end{figure}

\begin{figure}
\includegraphics[height=12cm,width=16cm]{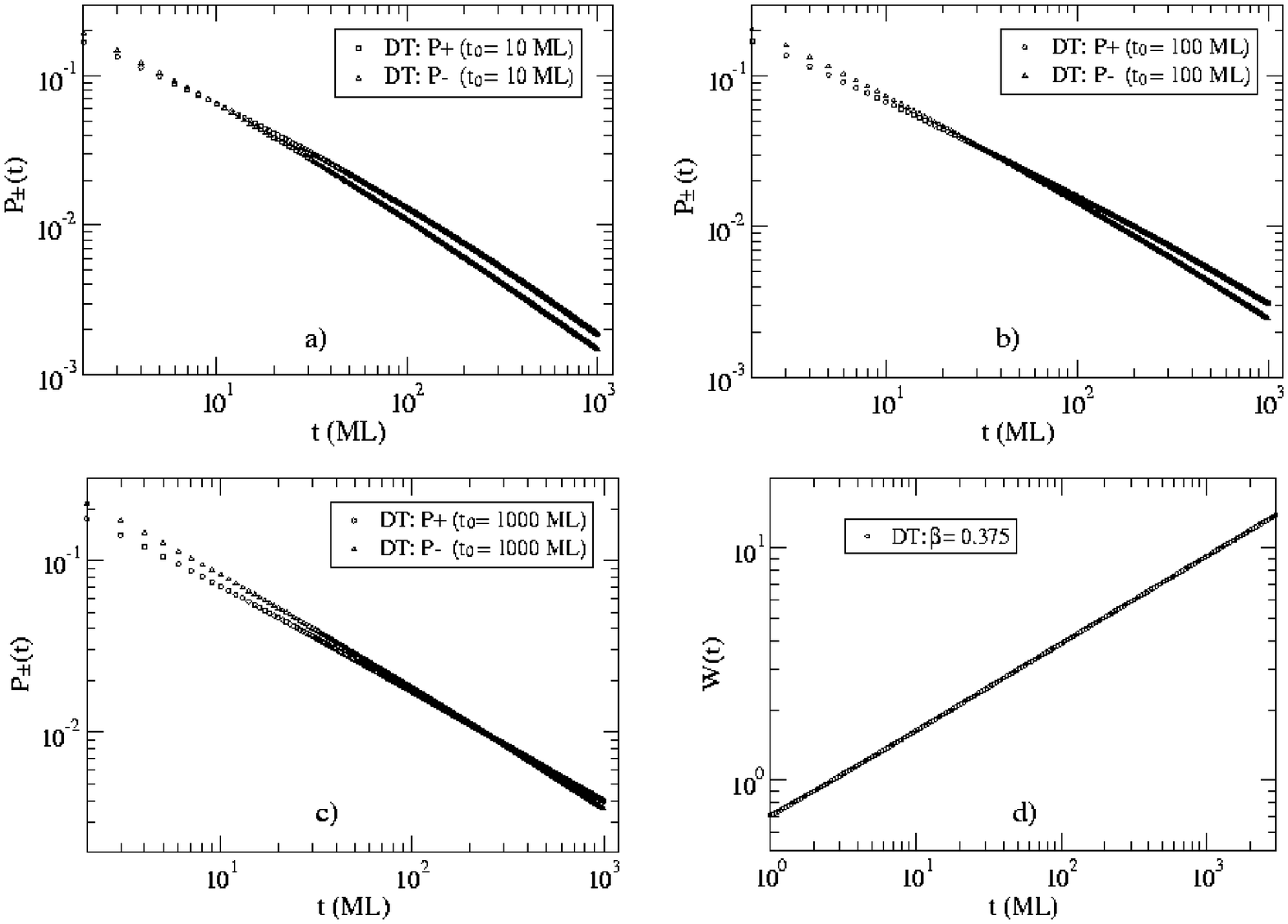}
\caption{\label{fig10} 
Positive and negative persistence probabilities [panels (a)--(c)]
for the DT model, obtained using different values of the initial 
time $t_0$. Persistence probabilities are computed starting from the 
configuration corresponding to: a) $t_{0}=10$ ML. As in 
the case of the F model, we do not find a clear power law for 
the persistence curves. b) $t_{0}=100$ ML; c) $t_{0}=1000$ ML: 
Power law decays are recovered and characterized by exponents that 
are approximately equal to those corresponding to the steady--state 
regime, $\theta^{S}_{+}\approx 0.64$ and 
$\theta^{S}_{-}\approx 0.71$. d) Log--log plot of the 
interface width $W$ as a function of time $t$ (in units of ML). 
The slope gives a growth exponent of $\beta_{DT} \simeq 0.375$.} 
\end{figure}

We recall that in Section \ref{sim1} the transient persistence 
exponents were measured from simulations in which the initial  
configuration was completely flat, corresponding to $t_0=0$. 
To examine the dependence of the persistence probabilities on 
the choice of $t_0$, we evolved samples governed by F and DT 
atomistic diffusion rules for $t_{0}=10$, $100$ and $1000$ ML, 
starting from perfectly flat initial states and used the 
resulting configurations as starting points for measuring the 
persistence probability (the probability of the height at a 
given site not returning to its initial value at time $t_0$) as a
function of $t$. We show the results of these simulations in 
Figs.~\ref{fig9} and \ref{fig10} for the F and DT models, 
respectively. We find that even for the small value of 
$t_0=10$ ML [see panel a)], the observed persistence probabilities 
do not exhibit power law decay in time with the transient 
persistence exponents, despite the fact that the expected 
condition~\cite{Krug1} for transient behavior, $t \gg t_{0}$, 
is well satisfied in a large part of the range of $t$ used in 
these simulations. These results point out a practical difficulty in
obtaining experimental evidence for transient persistence behavior.
Since perfectly flat initial configurations can hardly 
be achieved experimentally and experimental measurements 
are always started from a relatively rough substrate, the 
transient persistence exponents may very well not be measurable 
from experiments if the only way of measuring these exponents 
is to start from a perfectly flat morphology.

As the value of $t_0$ is increased to $100$ ML, the persistence 
probabilities tend to show the expected power law behavior, as 
shown in Fig.~\ref{fig9}(b) for the F model and in Fig.~\ref{fig10}(b) 
for the DT model. Most surprisingly, as shown in Figs.~\ref{fig9}(c) 
and \ref{fig10}(c), we find that for $t_{0}=1000$ ML, one recovers 
precisely the power law behavior, $P(t_{0},t_0+t) \propto 
t^{-\theta^{S}}$, and the exponents are essentially the same as 
the previously obtained  steady--state ones shown in 
Table~\ref{tab_short}. This investigation, thus, reveals the fact that 
a measurement of the steady--state persistence exponents does not
require the preparation of an initial state in the long--time steady--state
regime where the interface width has reached saturation: an initial state in 
the pre--asymptotic growth regime where the interface width is still 
increasing as a power law in time [as illustrated in 
Figs.~\ref{fig9}(d) and \ref{fig10}(d)] is sufficient for measurements of the
steady--state persistence exponents. A similar result was reported in 
Ref.~\cite{Krug1}, but it was argued there that the measurement time $t$
must be much smaller than $t_0$ for steady--state persistence behavior to
be observed. Our results show that the 
steady--state persistence exponents are found even if $t$ is 
of the order of (or even slightly larger than) the initial time 
$t_0$. This observation has an important practical benefit: it implies that 
one can easily obtain accurate estimates of the
steady--state persistence exponents using rather large systems ($L 
\sim 10^{4}$), and growing approximately up to $t_{0} \sim 10^{3}$ 
ML, instead of having to use the very large values (of the 
order of $ L^{z}$) of $t_0$ necessary for obtaining saturation of the 
interface width. At the same time, this observation also illustrates 
the above mentioned difficulty in obtaining the transient exponents
from experimental measurements.

\begin{figure}
\includegraphics[height=5.5cm,width=15.0cm]{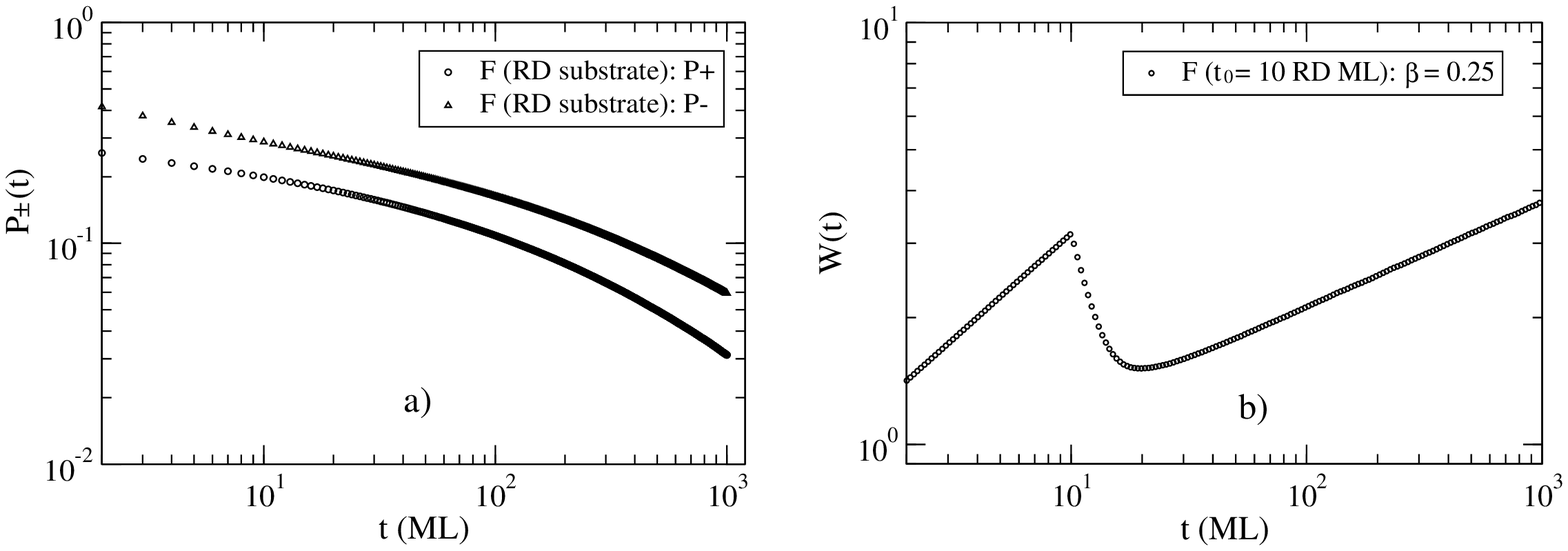}
\caption{\label{fig11}
a) Positive and negative persistence probabilities for the 
F model. During the deposition of the first 10 ML, 
the growth process is random deposition. The diffusion rules of the
F model are then used to evolve the interface.
Persistence probabilities are computed starting from the 
configuration obtained after the random deposition of 10 ML. 
The positive and negative 
persistence exponents in the last growth decade are in the range 
$0.6$ to $0.7$, depending on the fitting region. b) Log--log plot 
of the interface width $W$ as a function of $t$ (in ML). 
The slope in the first decade of $t$ is 
precisely the random--deposition value, $\beta=0.5$. 
The second decade shows a crossover 
region where the systems undergoes a transformation towards a 
morphology governed by the F model diffusion rules, and the last decade 
is characterized by the expected growth exponent of the F model, 
$\beta=0.25$.}
\end{figure}

\begin{figure}
\includegraphics[height=5.5cm,width=15.0cm]{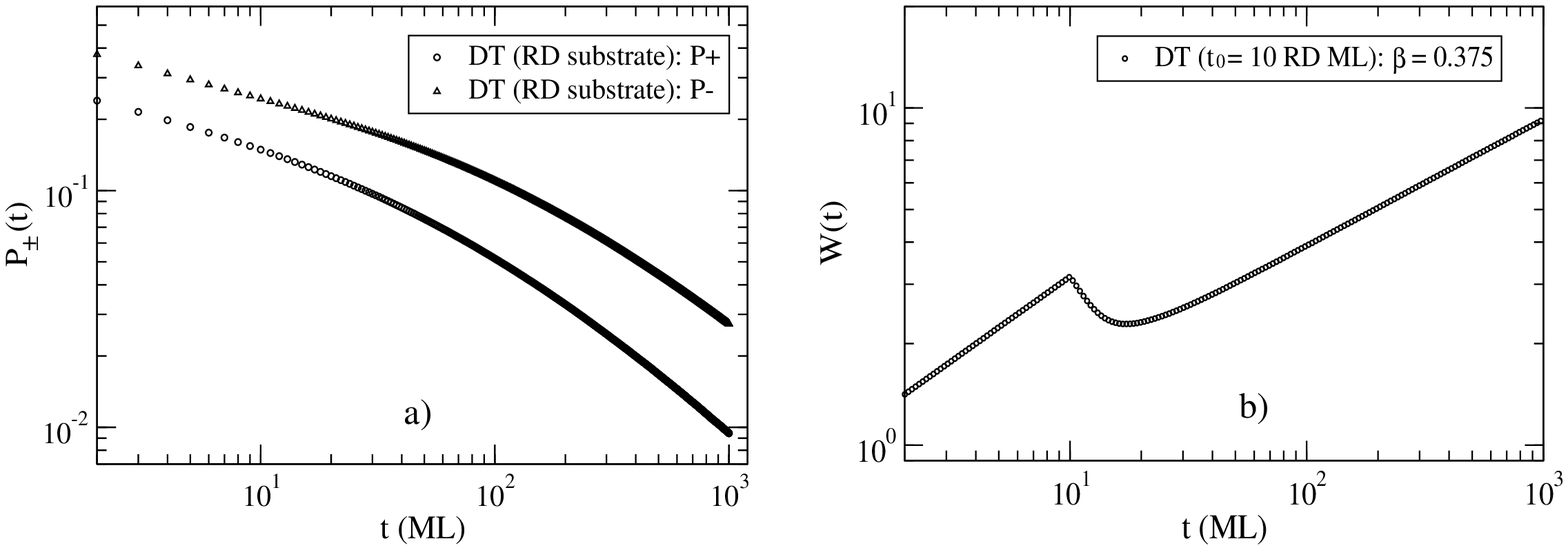}
\caption{\label{fig12} Positive and negative persistence 
probability curves for the DT model. During the deposition of the first 
10 ML, the growth process is random deposition. The diffusion rules of
the DT model are used to evolve the interface subsequently.
Persistence probabilities are computed starting from the 
configuration obtained after the deposition of the first 10 ML. 
The positive and negative 
persistence exponents in the last growth decade are 
approximatively equal to $0.66$ and $0.79$ respectively. b) 
Log--log plot of the interface width $W$ as a function of 
time $t$ (in ML). Beyond the crossover 
regime, the last decade in $t$ is characterized by the expected growth 
exponent of the DT model, $\beta \simeq 0.375$.}
\vspace{1.2cm}
\end{figure}

To investigate the effects of random imperfections in the 
initial substrate (which are always present in experimental studies)
on the persistence behavior, we carried out simulations
in which particles were deposited randomly on a perfectly 
flat substrate for 10 ML and the resulting configuration   
was used for further depositions using the diffusion rules of the
F and DT models. Persistence probabilities were calculated starting
from the configurations obtained after the random deposition of 10
ML. Figures \ref{fig11} and \ref{fig12} show the results for 
the F and DT models, respectively. We find that even when 
the persistence calculation starts from a configuration 
characterized by random deposition, there is an 
indication that one can still obtain the steady--state exponents 
during the last decade of $t$ where the growth exponent reaches 
the values characteristic of the diffusion rules of the 
specific (F or DT) model being considered. Indeed, in the time 
region where the growth exponents are $\beta=0.25$ for the 
F model [see Fig.~\ref{fig11}(b)] and $\beta=0.375$ 
for the DT model [see Fig.~\ref{fig12}(b)], we have calculated the 
persistence exponents and recovered values very close to 
the steady--state ones. These observations confirm our earlier conclusions
about the relatively easy measurability of the steady--state persistence  
exponents and the difficulty in measuring the transient exponents in
experimental situations.

\subsection{Persistence exponents in (2+1) dimensions}
\label{sim2}
Our calculations in (2+1)--dimensions make use of our observation 
(discussed above) concerning the possibility of obtaining the 
correct steady--state exponents from simulations that avoid the 
time consuming process of reaching the true steady state where 
the interface width has saturated. 
The result that the persistence exponents obtained from
(1+1)--dimensional simulations using fairly  small values of 
$t_{0}$ and $t \sim t_0$ are quite close to the steady--state
values allows us to extract numerically the steady--state 
persistence exponents in (2+1)--dimensions using 
systems with reasonably large sizes. If one had 
to run systems of size $L \sim 100 \times 100$ all the way to
saturation in order to measure the steady--state persistence 
exponents, it would have been impossible to do the calculations 
within reasonable simulation time. In addition, 
decreasing the system size is not an acceptable solution because 
the results then become dominated by finite size effects.

Simulations for (2+1)--dimensional discrete growth models 
were carried out for the F model ($\beta=0$) and the DT model 
($\beta \simeq 1/5$). Simulations using systems of size
$L=200 \times 200$ revealed that the growth exponents, 
obtained from averages over 200 independent runs, are 
$\beta=0.04 \pm 0.01$ and $0.20 \pm 0.01$ for the F and DT
models, respectively, in agreement with Ref.~\cite{WVuniv}. 
In the DT model, we noticed a crossover from the initial 
value of 0.26 to the asymptotic expected value of 0.20, 
indicating that no additional noise reduction technique 
is necessary for obtaining results that reflect the 
correct universality class of this model. For both F and 
DT models we calculate the transient and steady--state persistence 
probabilities by recording the fraction of sites which do not 
return to their initial height up to time $t$, as in the 
(1+1)--dimensional case. We used $t_0=0$ (perfectly flat initial state)
in the calculation of the transient persistence probabilities, 
and three different values, such as $t_{0}=20$ ML, $200$ ML and $2000$
ML for the F model, in the calculation of the steady--state exponents.

\begin{figure}
\includegraphics[height=7cm,width=10cm]{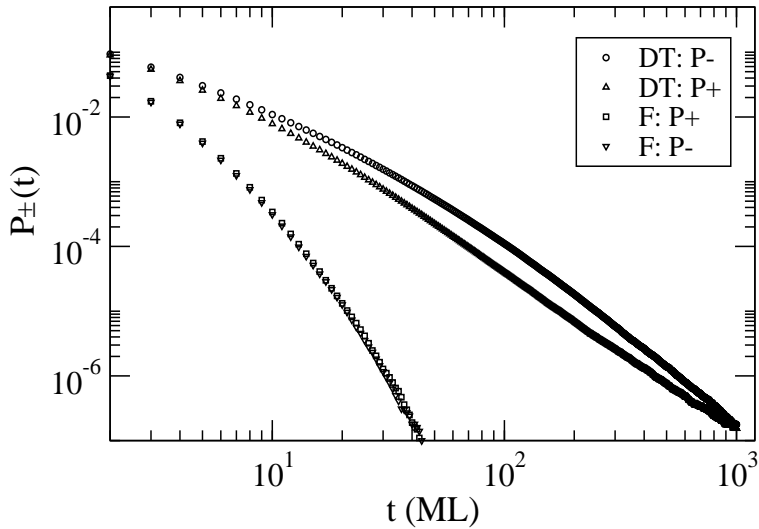} 
\caption{\label{fig13} 
Transient persistence probabilities  
for the (2+1)--dimensional F and DT growth models. In the case of the F 
model, systems of size $L=1000 \times 1000$ have been averaged 
over 200 independent runs, while for the DT model, systems of size 
$L=500 \times 500$ have been averaged over 800 independent runs. 
The transient persistence probability for the F model exhibits 
a very fast decay, characterized by a persistence exponent $\theta^{T} 
\approx 6.9$ for the last decade of $t$. More accurate results 
(see Table \ref{tab_2+1}) are obtained for the exponents in the DT model, 
although the statistics is not excellent.}
\end{figure}

We report the results for the transient probabilities just for 
the sake of completeness: the rapid decay of the persistence
probabilities prevents us from obtaining accurate values of
the associated persistence exponents. This fast decay of 
the transient persistence probability is a consequence of 
the reduced roughness of these higher dimensional models. 
This effect is particularly pronounced for the F model for 
which the persistence exponent is found to be larger than 
$6$ and the persistence probability decreases rapidly to 
zero for any deposition time larger than $\sim 60$ ML, 
as shown in Fig.~\ref{fig13}. We also observe that the 
transient values of the positive and negative persistence 
exponents in the DT model are roughly 3 times larger than 
the values obtained in the $(1+1)$--dimensional case. The 
relative difference between the positive and negative 
persistence exponents remains approximately the same as that in
the (1+1)--dimensional model. Our results for these 
(2+1)--dimensional persistence exponents are summarized 
in Table ~\ref{tab_2+1}.

\begin{table}
\begin{tabular}{|c|cc|cc|cc|cc|cc|cc|}
\hline \hline Growth model & $L$ && $\theta_{+}^{T}$ && $\theta_{-}^{T}$ &&
$\theta_{+}^{S}$ && $\theta_{-}^{S}$ && $\beta$ &
\\ \hline \hline FM & $200 \times 200$  && $> 6$ && $>6$ &&
$1.02 \pm 0.02$ && $1.00 \pm 0.02$ && $0.04 \pm 0.01$ &
\\ \hline DT & $200 \times 200$  && $2.84$ && $2.44$ &&
$0.76 \pm 0.02$ && $0.85 \pm 0.02$ && $0.20 \pm 0.01$ & \\ \hline \hline
\end{tabular}
%\vspace{0.5 cm} 
\caption{\label{tab_2+1} Transient and steady--state persistence 
exponents, $\theta_{\pm}$, for two (2+1)--dimensional discrete growth
models. The measured value of the growth exponent $\beta$ is shown in 
the last column. The transient persistence exponents are measured 
with relatively low accuracy due to the rapid temporal decay 
of the persistence probabilities.} 
%\vspace{1.0cm}
\end{table}

\begin{figure}
\includegraphics[height=12cm,width=16cm]{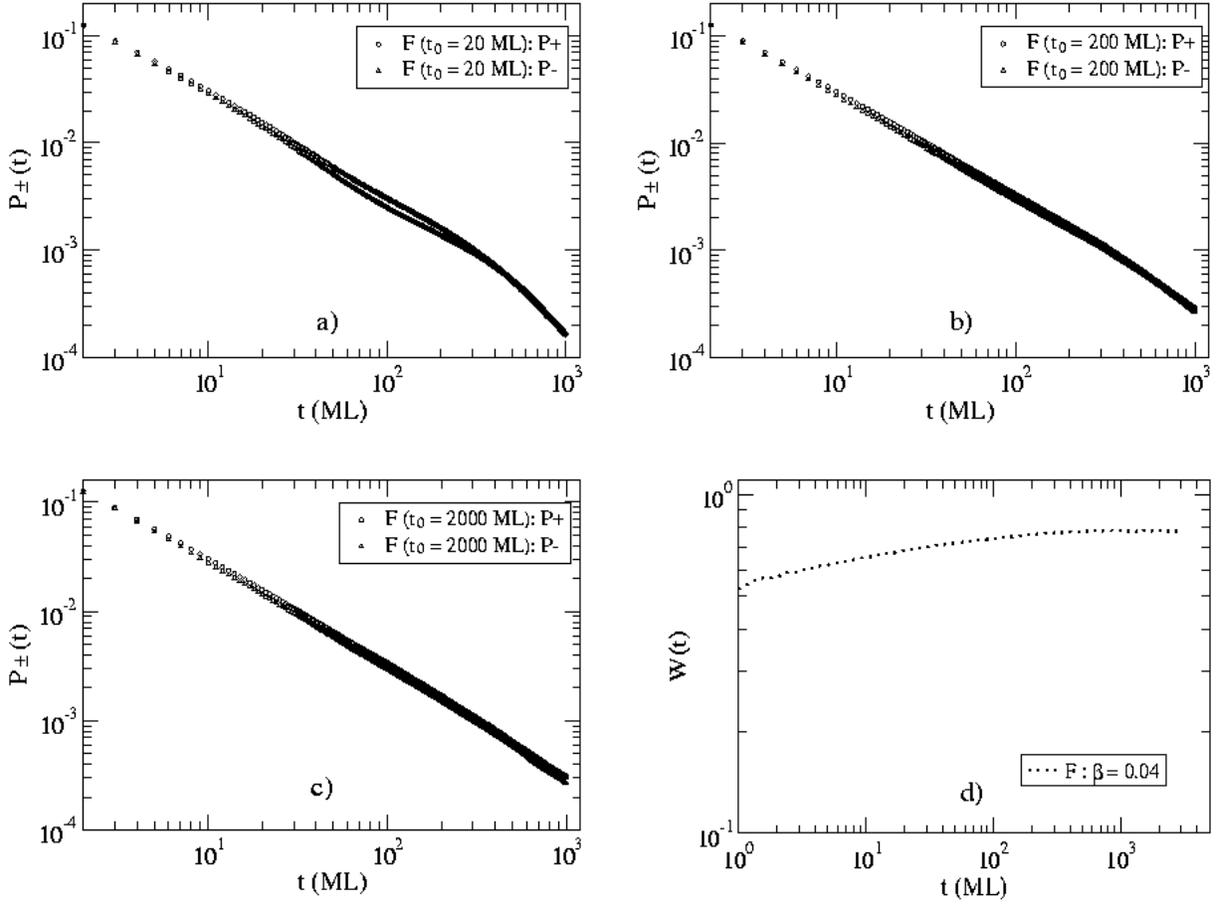}
\caption{
\label{fig14} 
Persistence probabilities for the (2+1)--dimensional F model 
of size $L=200 \times 200$, averaged over 200 runs, obtained 
from simulations with different values of the initial time $t_0$. 
a) $t_{0}=20$ ML. b) $t_{0}=200$ ML. c) $t_{0}=2000$ ML. The 
persistence probability curves in case c) show the expected 
power law decay characterized by the exponent values  
$\theta_{+}^{S}=1.02 \pm 0.02$ and 
$\theta_{-}^{S}=1.00 \pm 0.02$. d) Log--log plot of the 
interface width $W$ vs deposition time $t$ in units of ML. 
The slope in the intermediate growth decade is 
$\beta \simeq 0.04$ and thereafter it decreases to zero, 
as expected.}
\end{figure}

\begin{figure}
\includegraphics[height=12cm,width=16cm]{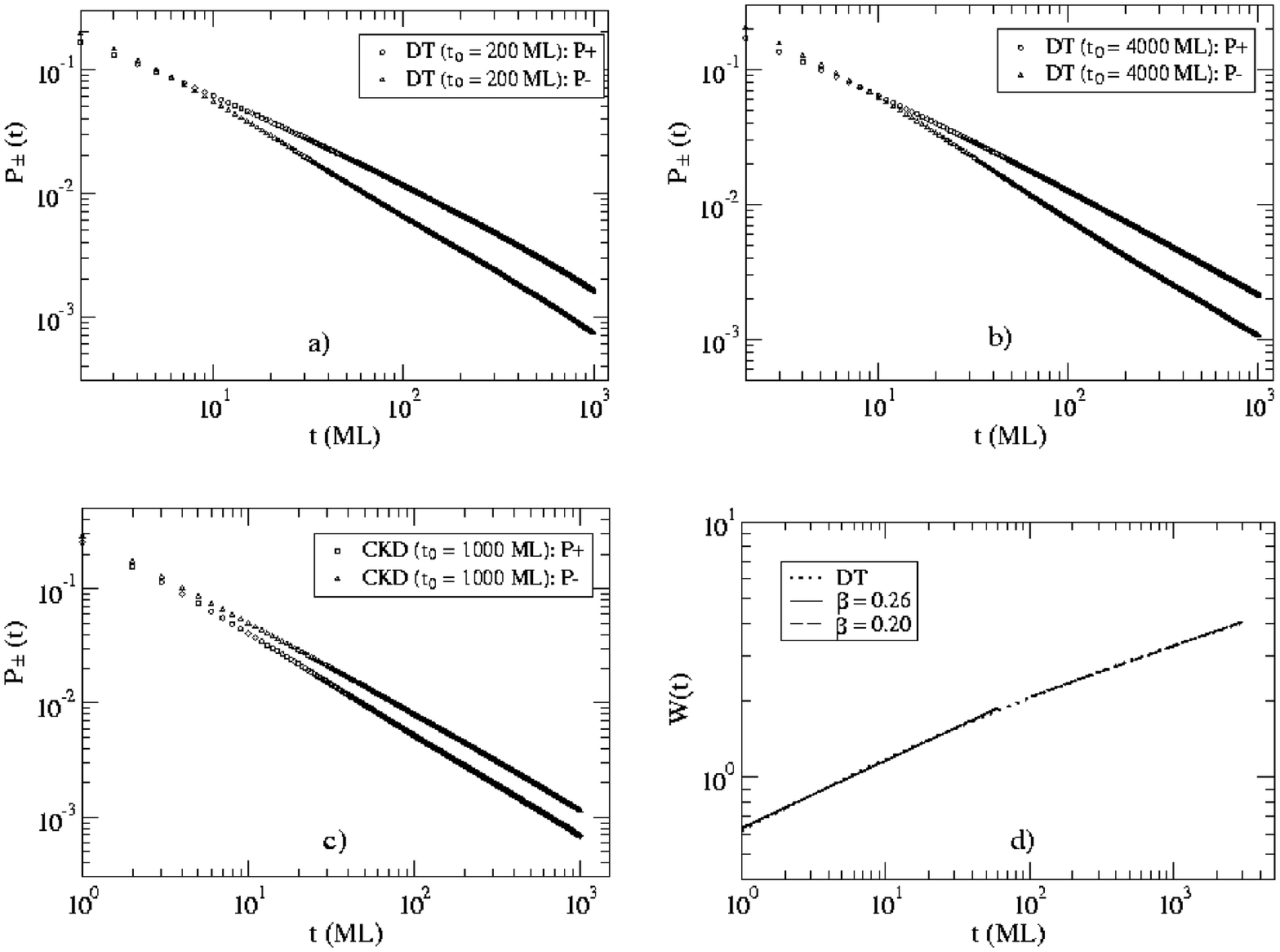}
\caption{
\label{fig15} 
Persistence probabilities for the 
(2+1)--dimensional DT model of size $L=200 \times 200$, averaged 
over 200 runs, obtained from simulations with different values of 
the initial time $t_0$. a) $t_{0}=200$ ML: The persistence 
probabilities do no exhibit clear power law decay. 
b) $t_{0}=4000$ ML: The persistence probability curves show 
the expected power law decay characterized by the exponent values
$\theta_{+}^{S}=0.77 \pm 0.02$ and 
$\theta_{-}^{S}=0.85 \pm 0.02$. c) Results for the 
(2+1)--dimensional CKD model, obtained using  
$L=100 \times 100$, $t_0=1000$ ML, $\lambda=5$ and $c=0.13$. 
d) Log--log plot of the interface width $W$ vs deposition 
time $t$ in ML for the DT model. The slope manifests a crossover from 
an initial value of $\sim 0.26$ to the asymptotic value of 0.20.}
\end{figure}

We now focus on the steady--state persistence exponents 
which, as discussed above, are found using relatively 
small values of $t_{0}$ and $t \approx t_0$. In 
Fig.~\ref{fig14}(a), we show that for $t_{0} \ll t$ (e.g. 
for $t_{0}=20$ ML), the persistence probability of the F 
model does not exhibit a clear power law decay. However 
panels b) and c) of Fig.~\ref{fig14} reveal that once 
$t_{0}$ becomes of the order of the measurement time $t$, 
the expected power law is recovered and in addition the 
steady--state exponent for the linear F model, 
$\theta^{S}=1.01 \pm 0.02$, which should be equal to 
$(1-\beta$) with $\beta=0$, is recovered. The results 
for the DT model are presented in Fig.~\ref{fig15}. 
The steady--state persistence exponents have been measured 
from the power law decays shown in Fig.~\ref{fig15}(b). 
In this temporal regime, as shown in panel d), the growth 
exponent is equal to the asymptotic value of $1/5$. 
The persistence behavior of the DT model in this regime 
is characterized by  $\theta^{S}_{+} \approx 0.76$ and 
$\theta^{S}_{-} \approx 0.85$, indicating that the relation of
Eq.~(\ref{sc1}) holds reasonably well for the (2+1)--
dimensional nonlinear MBE dynamics, 
%We also notice that the 
%positive exponent satisfies this relation better, 
as in the 
$(1+1)$--dimensional case. It is important to mention that 
the same values of the persistence exponents have 
been obtained using a DT system with $L=40 \times 40$, 
equilibrated for $t_0=10^5$ ML, as required in the traditional 
method (used in most of the (1+1)--dimensional simulations) 
of measuring the steady--state persistence probabilities. Thus the
``quick and easy'' method of obtaining the steady--state persistence
exponent again agrees well with the exponent extracted from the
actually saturated interface, as discussed above.

We have also performed some preliminary persistence calculations for the 
(2+1)--dimensional CKD model in order to check the validity of 
our reported (2+1)--dimensional MBE persistence exponents. Using 
$L=100 \times 100$ and $t_{0}=1000$ ML, we find that the values 
of the positive and negative persistence exponents 
depend to some extent on the chosen values for the 
coefficient $\lambda_{22}$ of the nonlinear term and the 
control parameter $c$. For example, we obtain  
$\theta^{S}_{+} \approx 0.82$ and  
$\theta^{S}_{-} \approx 0.77$ using $\lambda_{22}=5.0$ 
and $c=0.085$, and $\theta^{S}_{+} \approx 0.88$ and 
$\theta^{S}_{-} \approx 0.83$ using $\lambda_{22}=5.0$ 
and $c=0.13$. Both cases are characterized by a growth 
exponent of $0.18 \pm 0.01$, in agreement with 
Ref.~\cite{chandan_2dckd}, which is consistent with the 
expected value of $1/5$. The results obtained in the 
latter case are displayed in panel c) of Fig.~\ref{fig15} 
for the purpose of illustrating the similarity between the 
DT and CKD models. From these observations, we conclude 
that the (2+1)--dimensional DT and CKD persistence results 
are consistent with each other and they clearly reflect the 
nonlinearity of the MBE dynamical equation in the difference 
between the values of the positive and negative persistence 
exponents as expected for the up--down asymmetric generic nonlinear
situation. 

\subsection{Scaling behavior of the persistence probability}
\label{fss}

Since all the results described above have been obtained from simulations of
finite systems, it is important to address the question of how the 
persistence probabilities are affected by the finite system size. We have
already encountered such effects in our study of persistence probabilities for
(1+1)--dimensional models (see Table~\ref{tab_DT}), where it was found that
the measured values of the persistence exponents in the steady state 
increase slightly as the system size $L$ is decreased, while the value of
the growth exponent $\beta$ decreases with decreasing $L$. We did not 
investigate finite size effects in our study of the transient persistence
probabilities because these studies were carried out for large values of
$L$ and relatively small values of the time $t$. 

The qualitative dependence of the measured values of the steady--state
persistence exponents $\theta^S_{\pm}$ and the growth exponent $\beta$ on
the sample size $L$ is not difficult to understand. The steady--state
persistence probabilities $P^S_{\pm}(t_0,t_0+t)$ 
exhibit a power law decay with exponent $\theta^S_{\pm}$ as 
long as the time $t$ is small compared to the characteristic 
time scale $\tau(L)$ of the system which is proportional to $L^z$. The
decay of $P^S_{\pm}$ becomes faster as $t$ approaches and exceeds 
this characteristic time scale. Since this departure from 
power law behavior occurs at earlier times for smaller systems, 
the value of the persistence exponent extracted from a power law 
fit to the decay of the persistence probability over a fixed
time window is expected to increase as the system size is reduced. In a similar
way, the measured value of $\beta$ is expected to be smaller for relatively
small values of $L$ because the precursor to the saturation of the width at
long times occurs at shorter values of $t$ in smaller systems. Thus, the
general trends in the system size dependence of the persistence and growth
exponents are reasonable. However, it would be useful to obtain a more
quantitative description of these trends.

Since the characteristic time scale $\tau(L)$ 
(``equilibration'' or ``saturation'' time) of a system of linear 
size $L$ is proportional to $L^z$, one expects, in analogy with 
the theory of finite size scaling in equilibrium critical
phenomena, that the steady--state persistence probability 
$P_{\pm}^S(t)$ (in this discussion, we omit the initial time 
$t_0$ in the argument of $P_{\pm}^S$ because the steady--state
persistence probability is independent of the choice of $t_0$) would be a
function of the scaling variable $t/L^z$. Another time scale has to be taken
into consideration in a discussion of the scaling behavior of the persistence
probability. This is the sampling time $\delta t$ which is the time interval
between two successive measurements of the height at a fixed spatial point.
In our simulations of the atomistic growth models, the smallest value of
$\delta t$ is 1 ML because the heights are measured after each deposition of
one complete (ideal) monolayer. However, larger integral 
values of $\delta t$ can also be used in the calculation of the persistence
probabilities. Since experimental measurements are also carried out at
discrete time intervals, the presence of a finite value of $\delta t$ has to
be accounted for in the analysis of experimental data also. Note that
the persistence probability itself is mathematically defined,
$P(t_0,t_0+t)$, for continuous values of time $t$ whereas measurements
and simulations are necessarily done on discrete time.     

It has been pointed out in Ref.~\cite{deltat} that discrete--time sampling of
a continuous--time stochastic process does affect the measured persistence
probability. Such effects have been investigated in detail~\cite{p0}
in the context of a different stochastic probability (called the survival
probability in Ref.~\cite{p0}) that measures the probability of the 
interface height at a fixed position not returning to its time--averaged
value within time $t$. In that work, it was found that the survival 
probability measured for a system of size $L$ with sampling interval 
$\delta t$ is a function of the scaling variables $t/L^z$ and $\delta t/L^z$.
We expect a similar behavior for the steady--state persistence probabilities
measured in our simulations. Thus, the expected scaling behavior of 
$P_{\pm}^S(t,L,\delta t)$ is
\begin{equation}
P_{\pm}^S(t,L,\delta t) = f_{\pm}(t/L^z, \delta t/L^z),
\label{scaling}
\end{equation}
where the function $f_{\pm}(x_1,x_2)$ should decay as 
$x_1^{-\theta_{\pm}^S}$ for small
$x_1$ and $x_2 \ll 1$. 

\begin{figure}
\includegraphics[height=6.5cm,width=17cm]{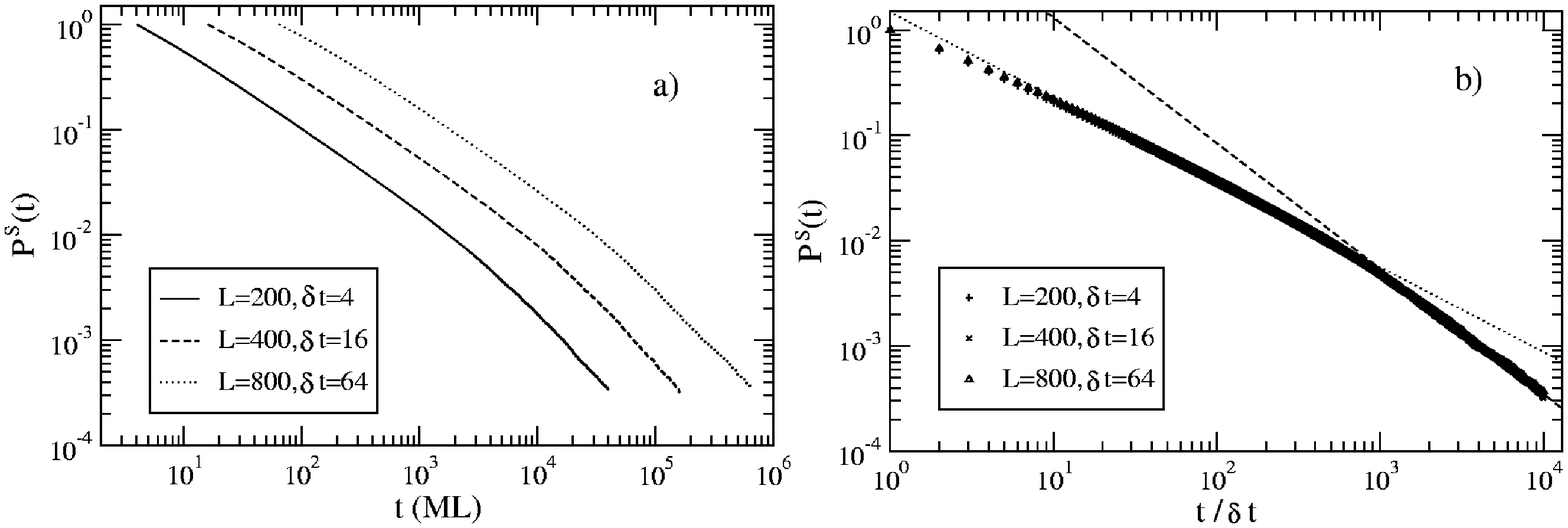}
\label{fig16}
\caption{Persistence probability, $P(t)$, for the F model shown for 
different system sizes with different sampling times. Panel a): Double--log plot
showing three different persistence curves vs. time corresponding to: 
$L=100$ and $\delta t=4$, $L=400$ and $\delta t=16$, $L=800$ and 
$\delta t=81$, respectively. Panel b): Finite size scaling of
$P(t,L,\delta t)$. Results for persistence probabilities for three
different sizes (as in panel a)) with the same value of $\delta t/L^z$
(i.e. $1/10^4$) are plotted versus $t/\delta t$ ($z=2$). The dotted (dashed) 
line is a fit of the data to a power law with an exponent of 
$\sim 0.75$ ($\sim 1.0$).}
\end{figure}

To test the validity of this scaling ansatz, we have carried out 
calculations of the steady--state persistence probability in the linear F
model (the positive and negative persistence probabilities are the same in
this model) using different values of $L$ and $\delta t$. Due to the
linearity of the F model, we have computed a persistence
probability $P^S(t)$ given by the average value of the positive and
negative persistence probabilities. If the scaling description of 
Eq.~(\ref{scaling}) is valid, then plots of $P^S(t,L,\delta t)$ 
versus $t/L^z$ for different values of $L$ and $\delta t$ should 
coincide if the value of $\delta t$ for the different sample sizes
are chosen such that the ratio $\delta t/L^z$ remains constant. As shown in
Fig. 16 where we present the data obtained from simulations of the 
(1+1)--dimensional F model for three different values ($200$, $400$ and $800$) of
$L$ and three correspondingly different values of $\delta t$ (4, 16 and 64,
so that $\delta t/L^z$ with $z=2$ is held fixed at $10^{-4}$), plots of
$P^S(t)$ versus $t/\delta t$ (which is proportional to $t/L^z$ because
$\delta t$ is chosen to be proportional to $L^z$) for the three different sample
sizes exhibit an excellent scaling collapse. These results confirm the validity
of the scaling form of Eq.~(\ref{scaling}).

As shown in Fig. 16, the scaling function $f$ exhibits 
the expected power law behavior for relatively small values of 
$t/\delta t$. Our results also show signatures of a crossover 
to a power law decay with exponent 1 as $t$ approaches and 
exceeds the characteristic time scale $\tau(L)$ (this crossover
occurs near $t/L^z \simeq 0.1$ in the F model). We discuss 
below a possible explanation for this behavior.

Height fluctuations at times $t_0$ and $t_0+t$ are expected to be completely
uncorrelated if $t$ is large compared to $\tau(L)$. Therefore, the persistence
probability $P^S(t)$ for values of $t$ much larger than $\tau(L)$ may be
obtained by considering a collection of fluctuating variables which have the
same probability distribution (since the system is in the steady state), and
which are completely uncorrelated with one another. Let $x_0, x_1, x_2, \ldots$
represent such a collection of variables (these variables may be thought of
as the height at a particular site measured at regularly spaced times with
spacing larger than $\tau(L)$). For simplicity, we assume that each 
$x_i$ is uniformly distributed between $-a$ and $a$. Then, given a particular
value of $x_0$, the probability $P_{+}(x_0,n)$ that all
the variables $x_i, 1\le i \le n$ are larger than $x_0$ may be easily obtained
as
\begin{equation}
P_{+}(x_0,n) = \left [\frac{1}{2a} \int_{x_0}^a dx\right ]^n = [(a-x_0)/(2a)]^n.
\label{calc1}
\end{equation}
The positive persistence probability $P_{+}(n)$ is obtained 
by averaging this probability over the probability distribution 
of $x_0$. Thus, we have
\begin{equation}
P_{+}(n) =\frac{1}{2a} \int_{-a}^{a} P_{+}(x_0,n) dx_0 = \frac{1}{n+1}, 
\label{calc2}
\end{equation}
which decays as a power law with exponent 1 for large $n$. 
This power law behavior does not depend on the form of the 
assumed probability distribution for the fluctuating variables 
$\{x_i\}$. Assuming a general probability distribution $p(x)$ 
with $\int_{-\infty}^\infty p(x) dx =1$, Eq.~(\ref{calc2})
can be written as
\begin{equation}
P_{+}(n) = \int_{-\infty}^\infty p(x_0) \left [\int_{x_0}^{\infty} p(x) dx 
\right ]^n dx_0 = \int_{-\infty}^\infty p(x_0) \left [1-\int_{-\infty}^{x_0}
p(x) dx \right ]^n dx_0.
\label{calc3}
\end{equation}
For large $n$, the quantity that multiplies $p(x_0)$ in Eq.~(\ref{calc3}) is of
order unity only for values of $x_0$ for which $\int_{-\infty}^{x_0} p(x) dx$
is of order $1/n$. Physically, this means that the positive persistence
probability is nonzero for large $n$ only if the initial value $x_0$ is very
close to the lower limit of the allowed range of values. This effectively 
restricts the upper limit of the integral over $x_0$
to $y_0$ where $y_0$ satisfies the requirement that
\begin{equation}
\int_{-\infty}^{y_0} p(x) dx = C/n,
\label{calc4}
\end{equation}
where $C$ is a constant of order unity. Since the quantity that multiplies
$p(x_0)$ in Eq.~(\ref{calc3}) is of order unity for such values 
of $x_0$, it follows that
\begin{equation}
P_{+}(n) \approx \int_{-\infty}^{y_0} p(x_0) dx_0 \propto \frac{1}{n}.
\label{calc5}
\end{equation}
This simple analysis shows that the simulation results for 
the behavior of the scaling function of Fig. 16 for 
large values of $t/\delta t$ are quite reasonable.

While we have not carried out similar scaling analyses
[Eq.~(\ref{scaling})] for other models, we
expect the scaling form of Eq.~(\ref{scaling}) to be valid in general.
We expect that such scaling analysis of the persistence probability as a 
function of the system size $L$ and the sampling time $\delta t$ would be
useful in the analysis of numerical and experimental data in the
future. In fact, we believe that a direct experimental verification of
the scaling ansatz defined by Eq.~(\ref{scaling}) will be valuable.

%====================================================================
\section{Concluding remarks}
\label{conclude} 
In this paper we have investigated the temporal first 
passage statistics, expressed in terms of temporal persistence 
probabilities, for a variety of atomistic models that provide 
discrete realizations of several linear and nonlinear Langevin 
equations for the stochastic dynamics of growing and fluctuating 
interfaces. Using extensive kinetic Monte Carlo simulations, 
we have obtained transient and steady--state persistence 
exponents for these (1+1) and (2+1)--dimensional SOS and 
RSOS growth models. We have followed the methodology of 
Krug {\it et al.} \cite{Krug1,Krug2} and extended their 
numerical work to the nonlinear MBE dynamical equation by 
studying the persistence behavior of the atomistic DT, WV, 
CKD and  KPK models. From these studies, we have identified 
two persistence exponents for each of the two temporal 
regimes (transient and steady--state) of the persistence 
probability. The difference between the values of the two 
exponents reflects the nonlinearity (and the resulting lack
of up--down symmetry) of the MBE dynamical equation.

Among the models studied here, we find that in (1+1) dimensions 
and in the range of system sizes used in our simulations, 
WV and DT models are hardly distinguishable from the point of 
view of transient and steady--state persistence behavior: the 
persistence exponents obtained for these two models are very 
close to each other. We, therefore, conclude 
that in the range of simulation parameters used in this 
study, the (1+1)--dimensional DT and WV models belong 
to the same universality class (namely the MBE universality 
class) as far as their persistence behavior is concerned. A 
separate investigation is required in order to understand the 
universality class of the WV model in (2+1) dimensions. The KPK model 
appears not to reflect well the nonlinear feature of the underlying 
dynamical equation in the values of the positive and negative 
persistence exponents. This is probably due to strong finite size 
effects arising from the small lattice sizes used in our traditional 
steady--state simulations (i.e. using $t_{0} \sim L^{z}$). These 
finite size effects appear to lead to overestimated persistence 
exponents [and underestimated growth exponent, consistent with 
Eq.~(\ref{theta_S})]. 

We have also investigated the CKD model, which is another discrete 
model belonging to the MBE universality class, our main goal being 
a closer examination of how the nonlinearity of the underlying continuum 
equation is reflected in values of the transient and 
steady--state persistence exponents. In this case we have 
obtained clearly different values for the positive and negative 
persistence exponents. 
The predictions of the CKD model concerning the persistence 
exponents have been checked by applying the noise reduction 
technique to the DT model. We found that for the MBE universality 
class, the steady--state persistence exponents in (1+1) dimensions are: 
$\theta_{+}^{S}=0.66 \pm 0.02$ and $\theta_{-}^{S}=0.78 
\pm 0.02$. These two values represent the average of the results obtained
for the CKD and the noise reduced DT models. These results suggest
that measurements of persistence exponents can be profitably used to
detect the presence of nonlinearity in the continuum 
equations underlying surface growth and fluctuation phenomena. 

The observed difference between the positive and negative steady--state
persistence exponents for the models in the MBE universality class
implies that the relation of Eq.~(\ref{theta_S}), known to 
be valid for linear Langevin equations (our results for the 
linear F and LC models are in agreement with this relation), 
can not be satisfied by both these exponents. Thus, it is clear 
that at least one of these steady--state persistence exponents is not
related to the usual dynamic scaling exponents in a simple way.
We have found that the relation of Eq.~(\ref{theta_S}) is approximately
satisfied (within the error bars of our numerical results) by the smaller
one of the two steady--state persistence exponents in
all the (1+1)-- and (2+1)--dimensional discrete stochastic growth 
models studied in this paper. We have also shown analytically that this relation
between the smaller persistence exponent and the growth exponent is,
in fact, exact. The smaller exponent
appears to correspond to the positive (negative) persistence probability
if the top (bottom) part of the steady--state interface profile is smoother.
This observation suggests a deep (and potentially important) 
connection between the surface morphology and the associated 
persistence exponent, which has no simple analog in the dynamic scaling
approach where the critical exponents 
($\alpha$, $\beta$, $z=\alpha/\beta$) by themselves do 
not provide any information
about the up--down symmetry breaking in the surface morphology. 
Further investigation of this aspect would be very interesting and
highly desirable, particularly if experimental information on
persistence properties of nonequilibrium surface growth kinetics
becomes available.  

Our investigation of the effects of the initial configuration on the 
persistence probabilities indicates that the transient persistence 
exponents can be obtained only if the interface is completely flat at
the initial time. This restriction puts severe
limits on the possibility of measuring the transient persistence 
exponents in real experiments where it would be very difficult, 
if not impossible, to meet the requirement of zero initial roughness. 
We have also found the surprising and useful result that the steady--state
persistence exponents can be accurately measured even if the initial
configuration corresponds to a value of $t_0$ that is much smaller than
the time required for the interface to reach saturation. In other words,
the persistence probabilities exhibit their steady--state behavior for
measurement times comparable to the initial time $t_0$ even if the value
of $t_0$ is much smaller than $L^z$. This behavior was found in both (1+1)
and (2+1) dimensions, in all the linear and nonlinear models we
studied. This finding is very useful because it opens up 
the possibility of numerically calculating the steady--state 
persistence exponents for large systems and for higher dimensions as well. 
In fact, this observation enabled us to calculate the steady--state 
persistence exponents for (2+1)--dimensional models belonging to 
the EW and MBE universality classes. For the MBE universality class, we have 
considered the DT model and found the positive and negative 
persistence exponents in the steady--state to be $\approx 0.76$ 
and $\approx 0.85$, respectively, in (2+1) dimensions. 

We have also examined in detail the dependence of the measured steady--state
persistence probability in the (1+1)--dimensional F model on the sample
size $L$ and the sampling interval $\delta t$ which is always finite in
simulations and experimental measurements. We found that this dependence
is described by a simple scaling form. The scaling function was found to 
exhibit power law decay with with exponent 1 for times larger than $L^z$.
We have proposed a simple explanation for this behavior. We believe that such
scaling analysis would prove to be useful in future numerical and experimental
studies of persistence properties.

We conclude from the results of this study that persistence 
probabilities provide a valuable set of tool for investigating 
the dynamics of nonequilibrium systems in general, and surface 
growth and fluctuations in particular. Recent experimental studies 
have shown that the concept of persistence can be applied to analyze 
the dynamics of fluctuating steps on Al/Si(111), Ag(111) and Pb(111) 
surfaces \cite{Dan,alex} recorded using STM methods. We believe 
that in view of the importance of thermal and shot--noise fluctuations 
in the dynamics of growing and  fluctuating interfaces, theoretical 
and experimental studies of persistence would play an important role 
in the analysis of the dynamics of nonequilibrium surface growth.

\vspace{0.5 cm}
\begin{center}
{\bf ACKNOWLEDGMENTS}
\end {center}
We thank E. D. Williams and D. B. Dougherty for discussions. 
This work was partially supported by the US--ONR, the LPS, and the
NSF--DMR--MRSEC.
%====================================================================

%==================================================================

\begin{thebibliography}{99}
\bibitem{note} The asymptotic long--time behavior of the (1+1)--dimensional
Wolf--Villain model is known to be in the Edwards--Wilkinson universality
class. However, its behavior in the relatively short--time regime considered
here is expected to be similar to that of other models in the MBE universality
class.
\end{thebibliography}
\end{document}